\documentclass[preprint, 12pt]{elsarticle}

\usepackage[]{graphicx}
\usepackage[]{verbatim}
\usepackage[dvipsnames]{xcolor}
\usepackage{tabularx}

\usepackage{amsfonts}
\usepackage{amsmath}
\usepackage{amssymb}
\usepackage{bm}
\usepackage{graphicx}
\usepackage[breaklinks,colorlinks=true,citecolor=blue]{hyperref}
\usepackage[]{cleveref}
\usepackage{mathrsfs}
\usepackage[lofdepth,lotdepth,caption=false]{subfig}
\usepackage{varwidth}
\usepackage{wrapfig}
\usepackage{times}
\usepackage{longtable}
\usepackage{multirow}
\usepackage{tikz}

\usepackage[margin=2.5cm]{geometry}
\journal{Current Applied Physics}
\begin{document}
\begin{frontmatter}

\title{
  Epitaxially strained ultrathin LaNiO$_3$/LaAlO$_3$ and 
  LaNiO$_3$/SrTiO$_3$ superlattices:\\
  a density functional theory + $U$ study
}

\author[mymainaddress,mysecondaryaddress]{Heung-Sik Kim\corref{mycorrespondingauthor}}
\ead{heungsikim@kangwon.ac.kr}
%\email{heungsikim@kangwon.ac.kr}
%\affiliation{Department of Physics, Korean Advanced Institute of Science and Technology, Daejun 305-701, Korea}
%\affiliation{Department of Physics and Institute of Quantum Convergence Technology, Kangwon National University, Chuncheon 24311, Republic of Korea}

\author[mymainaddress]{Sang Hyeon Park}
%\affiliation{Department of Physics, Korean Advanced Institute of Science and Technology, Daejun 305-701, Korea}

\author[mymainaddress]{Myung Joon Han\corref{mycorrespondingauthor}}
\ead{mj.han@kaist.ac.kr}
%\affiliation{Department of Physics, Korean Advanced Institute of Science and Technology, Daejun 305-701, Korea}

\address[mymainaddress]{Department of Physics, Korea Advanced Institute of Science and Technology,
Daejun 305-701, Korea}
\address[mysecondaryaddress]{Department of Physics and Institute of Quantum Convergence Technology, Kangwon National University, Chuncheon 24341, Korea}
\cortext[mycorrespondingauthor]{Corresponding authors}

\begin{abstract}
By employing first-principles electronic structure calculations we investigate nickelate superlattices [LaNiO$_3$]$_1$/[LaAlO$_3$]$_1$ and [LaNiO$_3$]$_1$/[SrTiO$_3$]$_1$ with (001) orientation under epitaxial tensile strain. Within density functional theory augmented by mean-field treatement of on-site electronic correlations, the ground states show remarkable dependence on the correlation strength and the strain. In the weakly and intermediately correlated regimes with small epitaxial strain, the  charge-disproportionated insulating states with antiferromagneitc order is favored over the other orbital and spin ordered phases. On the other hand, in the strongly correlated regime or under the large tensile strain, ferromagnetic spin states with Jahn-Teller orbital order become most stable. The effect from polar interfaces in LaNiO$_3$]$_1$/[SrTiO$_3$]$_1$ is found to be noticeable in our single-layered geometry. Detailed discussion is presented in comparison with previous experimental and theoretical studies.
\end{abstract}


\end{frontmatter}

\section{Introduction}

Rare-earth nickelates have been attracting significant research interests by exhibiting a number of distinct phases including
non-Fermi liquid, charge-disproportionated (CD) insulating states with and without magnetic orders, and unconventional superconductivity more recently \cite{RNOPD,RNOreview,RNOh_review,Catalano2018_Review, Li2019Nature,Nomura2022rev,Sakakibara2020,Jiang2019,Nomura2019,Ryee2020}.
In $R$NiO$_3$ ($R$ = rare-earth elements) and their heterostructures, the coexistence of competing charge, spin, orbital, and 
lattice degrees of freedom (DoF) is known to host rich electronic behavior, thereby  providing a great chance to engineer the materials with desired
functional properties \cite{Catalano2018_Review}. Recent advancement in the atomistic thin-film growth technique, for example, enables the
application of epitaxial lattice strain and quantum confinement effect to various
transition metal oxides. 
Among them LaNiO$_3$ (LNO), a paramagnetic metal showing a normal Fermi-liquid behavior in its bulk phase, 
has been a focus of such efforts \cite{RNOh_review}. It was theoretically suggested 
that a cuprate-like Fermi surface and antiferromagnetism can be realized 
in LNO/LaAlO$_3$ (LNO/LAO) (001)-superlattice \cite{Hansmann2009PRL}. It was followed by subsequent studies on the related systems, demonstrating that LNO films or superlattices are insulating and 
magnetic in the ultrathin limit while their detailed natures are still not quite clearly understood \cite{Boris:2011vt,Gray2011PRB,King2014NNT,Golalikhani2018,Han:2010cy,Han:2011jl,MJH2011PRL,Han:2012bc,Han:2014be,Han:2014be,Yoo2016PRB}.
These observations signified the role of heterostructure effect and electron correlation in LNO, and motivated the later studies on so-called orbital engineering.

A number of {\it ab-initio} theoretical approaches have been employed to investigate these nickelates, including density functional theory (DFT), DFT+$U$, and more recently DFT augmented by dynamical mean-field theory (DFT+DMFT) \cite{Paul2019_Review,Han:2010cy,Han:2011jl,MJH2011PRL,Han:2012bc,Han:2014be,Han:2014be}. While they provide key insight and information, a common limitation of using small size of unicell is clearly noticed. It can possibly hamper the detailed comparison with experiments. There was one study of self-interaction corrected DFT that takes a $2\sqrt{2} \times 2\sqrt{2}$ supercell. Notably, this calculation well reproduces the antiferromagnetic (AF) order accompanied by CD although this solution is only obtained with the significantly large strain value of +4\% compared to bulk LNO \cite{Puggioni2012PRB,Boris:2011vt,Frano2013PRL}. For DFT+$U$ method, on the other hand, the larger supercell has not been employed, and therefore the often-obtained and seemingly-incorrect ferromagnetic (FM) ground state could not be properly assessed. %Therefore, despite the limitations of a mean-field approximation, DFT+$U$ studies with a large unit cell can give useful insights about possible ground state of ultrathin LNO systems and their  underlying physics.

In this work, we carried out extensive DFT$+U$ calculations of ultrathin LNO 
(001)-superlattices (SL), and investigated the effect of on-site Coulomb interaction and epitaxial strain. We considered two different SL types of LNO$_1$/LAO$_1$ and LNO$_1$/STO$_1$; {\it i.e.}, LNO layer is alternating with LaAlO$_3$ (LAO) and SrTiO$_3$ (STO). Contrary to the case of LNO/LAO, LNO/STO is nominally expected to have
the internal electric field. Thus the comparive study can give useful insight for the 
interfacial effect possibly occuring in the experimental situations\cite{Kaiser2011PRL,Tung2017,Geisler2017}. It is found that both SL systems favor AF-CD insulating phase in the weak-to-intermediate interaction regime ($U_{\rm eff} = 2 \sim 3$ eV) and under small tensile strain. It corresponds to the solution of self-interaction-corrected DFT 
but has not been obtained before within DFT+$U$. 
When tensile strain becomes larger, contrary to the simple expection of the enhanced $d_{x^2-y^2}$  occupation and the orbital polarization accompanied by AF order, a Jahn-Teller (JT) distorted antiferro-orbital order is stabilized with FM spin order. This FM-JT phase is also found to be the ground state in the large interaction regime of $U_{\rm eff} > 3$ eV regardless of strain. The comparison of two SL systems shows at most quantitative differences.

The paper is organized as follows: After describing the computation details in Sec.\ref{sec:details}, we present the overall phase diagram in Sec.\ref{sec:PD}. The detailed discussion of electronic structure and other underlying physics are given in Sec.\ref{sec:LAO} and \ref{sec:STO}. Discussion and further comparison with previous experimental and theoretical works are provided in Sec.\ref{sed:discussion}.

\section{Computational details and initial trial configurations}
\label{sec:details}
Vienna {\it ab initio} simulation package\cite{VASP1,VASP2} was employed for
DFT calculations. A 500 eV of energy cutoff is used for the projector-augmented wave (PAW)
 basis set. $4\times 4\times 3$, $6\times 6\times 3$ 
and $8\times 8\times 3$ {\it k}-point samplings on the Monkhorst-Pack grid are chosen 
for $2\sqrt{2} \times 2\sqrt{2}$, $2 \times 2$, and $\sqrt{2} \times \sqrt{2}$ supercells, respectively.
On-site electron correlations for Ni $d$ orbitals are treated within the simplified rotationally invariant DFT+$U$ method suggested by Dudarev {\it et al.}\cite{Dudarev1998PRB}. While there has been significant progress recently on understanding the functional forms and their differences \cite{HChen2016,HChen2016b,HPark2015,Ryee2018b,Ryee2018,Jang2018,Jang2021,Lee2019}, we chose this particular prescription because it was widely used in the previous nickelate studies. The effective on-site Coulomb interaction parameter $U_{\rm eff}\equiv U - J$ for Ni $d$
is varied from 0 to 5 eV. The inclusion of $U_{\rm eff}$ for La $f$ orbital makes no significant change except for the rather trivial upward shift of the unoccupied La $f$ states.

\begin{figure}
  \centering
  \includegraphics[width=0.45\textwidth]{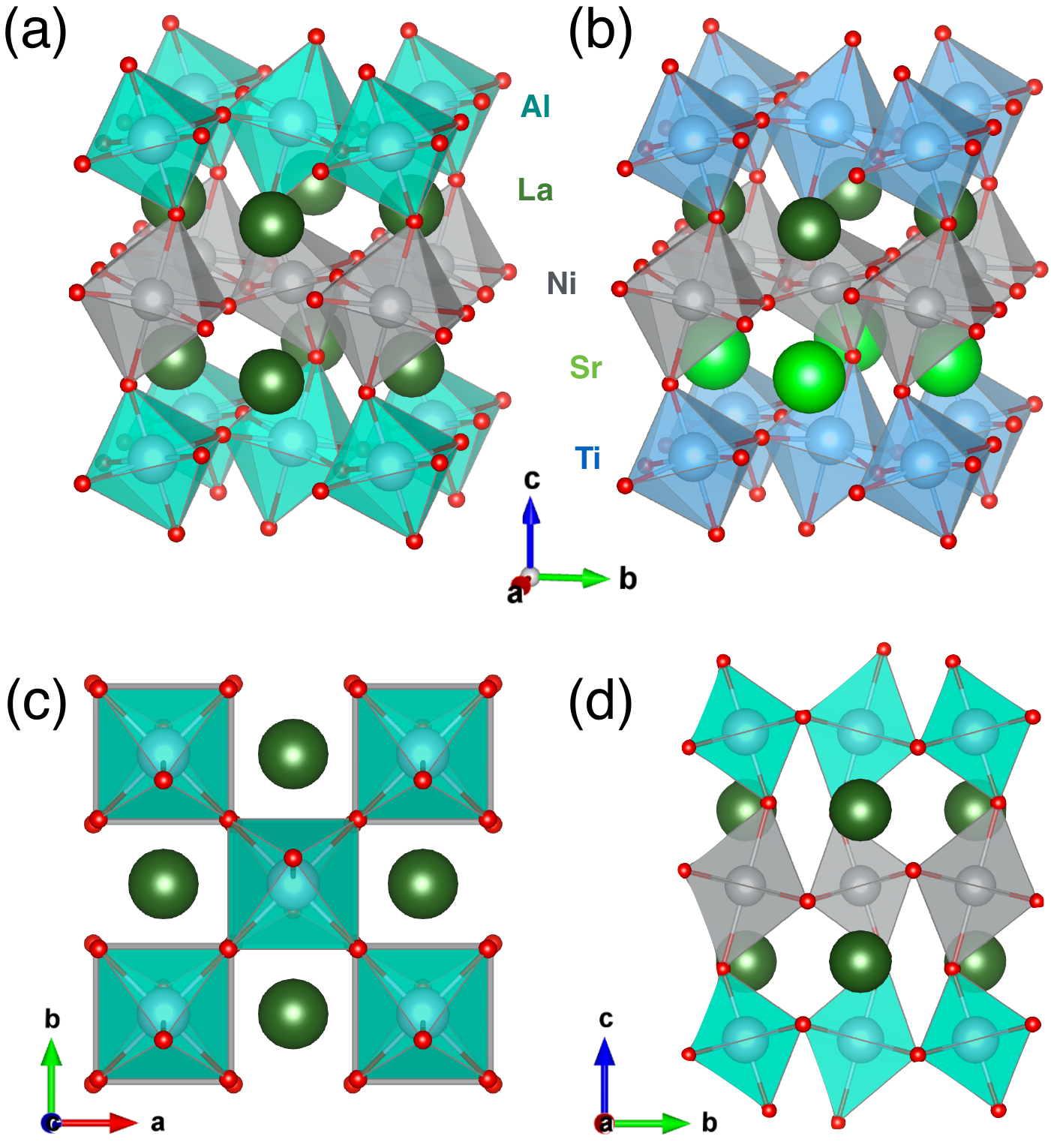}
  \caption{
  Stacking configurations of LNO and LAO/STO layers in (a) LNO$_1$/LAO$_1$ and (b) LNO$_1$/STO$_1$ superlattices adopted in the calculations. 
  In LNO/STO, asymmetric electrical potential is exerted on NiO$_2$ layer by the imbalance of La$^{3+}$ and Sr$^{2+}$ cations. (c, d) The ($a^-a^-c^0$)-type rotation of the NiO$_6$ octahedra in LNO/LAO SL.
  }
  \label{fig:struct}
\end{figure}

Fig.~\ref{fig:struct}(a) and (b) shows the unit cell geometry of LNO$_1$/LAO$_1$  
and LNO$_1$/STO$_1$ considered in this study, respectively. 
To study the effect of tensile strain
we consider the in-plane lattice constants $a$ = 3.91, 4.02, 4.10 \AA~ which 
correspond to 1.02, 1.05, 1.07 times of bulk LNO lattice constant, respectively. The first value is the experimental lattice constat of STO. For each configuration with different parameters ({\it i.e.}, in-plane lattice constant, substrate, and $U_{\rm eff}$ values), the out-of-plane $\hat{c}$-axis lattice constant and the
internal coordinates are optimized with a force criterion of 1 meV / \AA~. A octahedral rotation pattern depicted in Fig.~\ref{fig:struct}(c) and (d), namely, ($a^-a^-c^0$)-type in Glazer notation, has been considered in the currnt study
as it was found to be most stable in the perovskite heterostructures under tensile strain \cite{May2010PRB,Rondinelli2010PRB,Hatt2010PRB,Rondinelli2011PRL}.

\begin{figure*}
  \centering
  \includegraphics[width=0.95\textwidth]{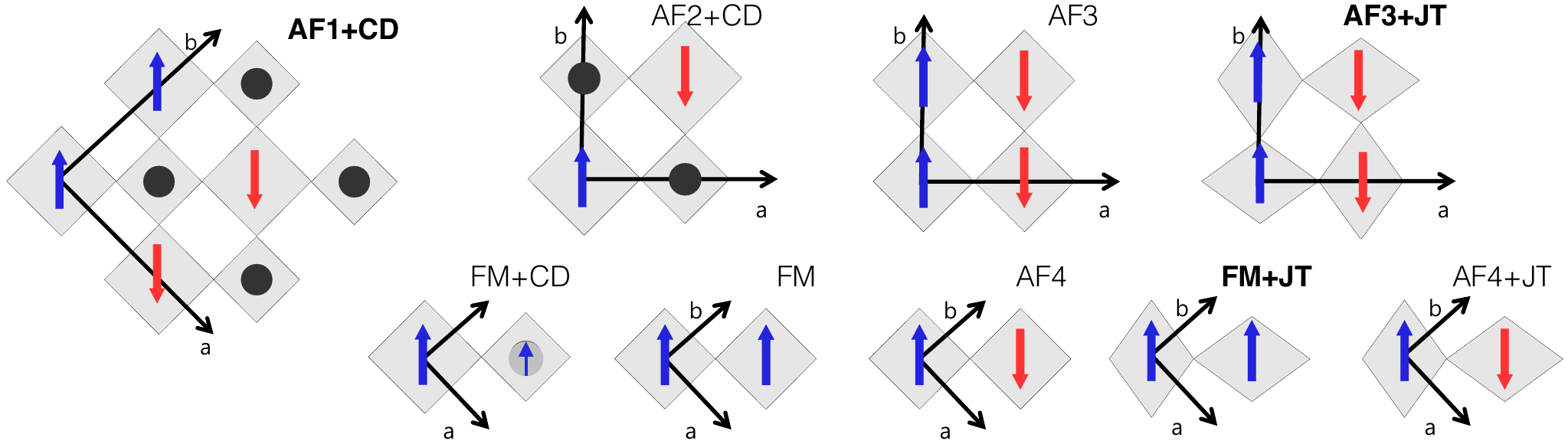}
  \caption{(Color online)
  Schematic diagrams that depict the NiO$_6$ octahedral distortion and the magnetic orders obtained from our calculations. The rotation of octahedra is not illustrated for simplicity. The magnetic (Ni1) and non-magnetic Ni (Ni2) is represented by the colored (blue and red) arrows and the black-filled circle, respectively. Among the converged solutions at the given structural configurations, the ground states are denoted as the thick labels (`AF1+CD', `AF3+JT', and `FM+JT'). 
  Note that, in the FM+CD phase, the moment at the Ni2 site is not zero. 
  }
  \label{fig:unitcells}
\end{figure*}

In order to obtain the ground state configurations,
we investigate an exhaustive set of initial unit cells up to $2\sqrt{2}\times 2\sqrt{2}$ in-plane periodicity (containing 8 Ni sites per a LNO layer) with which one can simulate JT and CD structural phases together with the corresponding spin and orbital orders. Fig.~\ref{fig:unitcells} summarizes the converged calculation results where the grey diamonds represent the shape of NiO$_6$ octahedra, and the colored arrows depict the direction of spins. The filled black circles refer to the non-magnetic Ni sites. 
Hereafter we denote the magnetically active and inactive Ni sites as Ni1 and Ni2, respectively.
As depicted in Fig.~\ref{fig:unitcells}, the oxygen octahedral cage surrounding Ni1 is larger than that of Ni2. Note that, in FM+CD phase, the magnetic moment at the Ni2 site
is not completely vanishing. 
Depending on the presence of CD and/or JT distortions and the type of AF order, 
six distinct AF phases are obtained;
CD with stripy (AF1+CD) or checkerboard AF (AF2+CD) order, 
JT with stripy (AF3+JT) or checkerboard AF (AF4+JT) order, 
and AF3/AF4 phases without any octahedral distortions. 
There are also three distinct FM order; FM, FM+CD, and FM+JT. 
Among them, AF1+CD and FM+JT order are found to be the ground states in the $U$-strain space as shown in Fig. \ref{fig:PD}.

\section{Phase diagram in $U$--strain space}
\label{sec:PD}

\begin{figure*}
  \centering
  \includegraphics[width=1.0\textwidth]{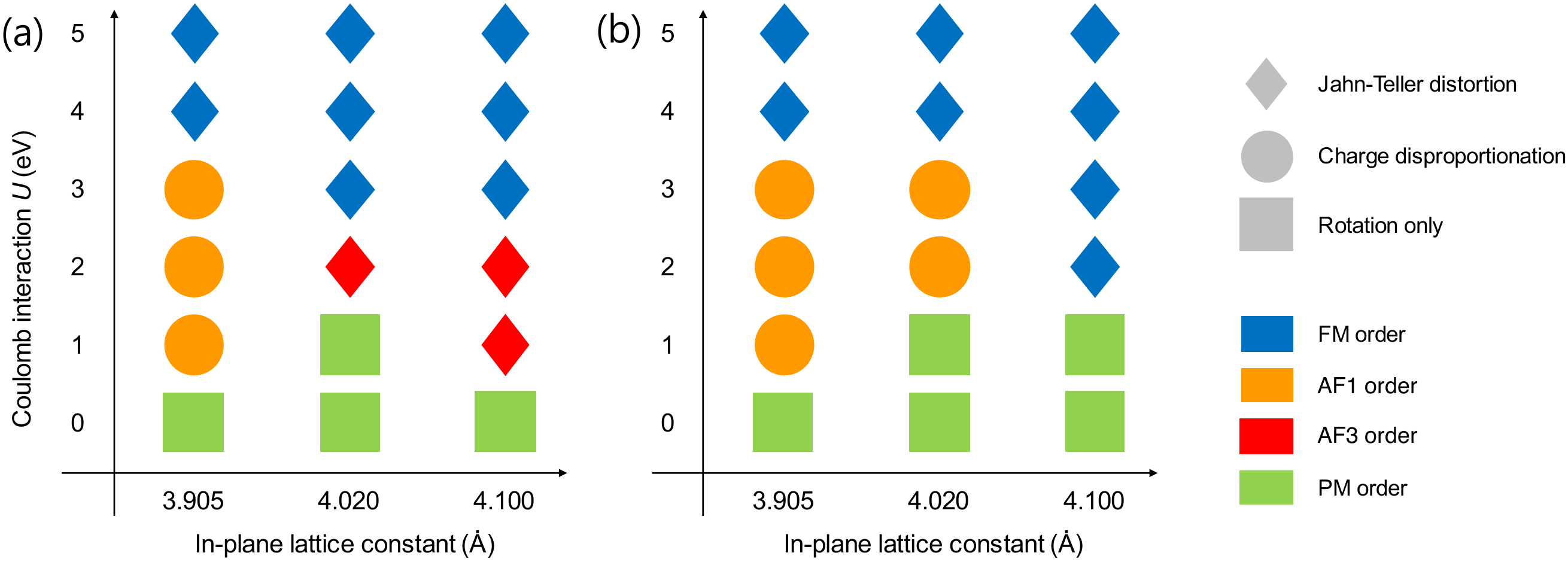}
  \caption{(Color online)
  Phase diagrams in the strain-$U_{\rm eff}$ parameter space for
  (a) LNO/LAO and (b) LNO/STO SLs. Colors
  and shapes of the symbols show the types of magnetic orders and the
  NiO$_6$ cage distortions, respectively. 
  }
  \label{fig:PD}
\end{figure*}

Fig.~\ref{fig:PD} presents the calculated phase diagram of LNO/LAO and LNO/STO SLs in the strain--$U_{\rm eff}$ parameter space. The ground state configuration of AF1+CD and FM+JT is depicted by the orange circle and the blue diamond, respectively. It is noted that the AF1+CD phase is stabilized in the regime of intermediate $U_{\rm eff}$ and small tensile strain for both SL systems while the FM+JT  becomes most stable as $U_{\rm eff}$ or strain increases. In the sense that both SLs are insulating in the reasonable  $U_{\rm eff}$ regime of $U_{\rm eff} \geq$ 3 eV, our result is consistent with the experimental observation of the insulating behavior in the ultrathin LNO 
SLs \cite{Boris:2011vt,Kaiser2011PRL,Gray2011PRB,Frano2013PRL,King2014NNT}. 
It is interesting to note that the AF1+CD ground state was reported for tensile-strained LNO/LAO with $a$ = 4.02\AA~ in a recent self-interaction-corrected LDA \cite{Puggioni2012PRB} and a slave-rotor mean-field study \cite{Bayo2013PRL}, but has not been obtained in DFT+$U$. Our result shows that it is not attributed to the different theoretical approximations, but simply to the different unit cells adopted in the previous studies \cite{Pentcheva2011PRB,Park2012PRL}.

We pay special attention to the following common features found for both LNO/LAO   (Fig.~\ref{fig:PD}(a)) and LNO/STO (Fig.~\ref{fig:PD}(b)).
(i) The phase diagrams can be roughly divided into the weakly, 
intermediately, and strongly correlated regime. It is also consistent with a previous  DFT+$U$ result \cite{HSK2015PRB}.
(ii) The insulating phase with AF order ({\it i.e}, AF1+CD or AF3+JT) is observed in the intermediate $U_{\rm eff}$  regime with small tensile strain. It is more relevant than FM+JT 
phase to the experimentally observed insulating phase with no net magnetization  \cite{Boris:2011vt,Frano2013PRL}.
(iii) Tensile strain stabilizes the FM+JT phase, which is contrary to the expectation of Ref.~\cite{Chaloupka2008PRL} which 
predicts the $d_{x^2-y^2}$ orbital order with AF spin order under tensile strain. Note that this tensile-strain-induced
JT phase was also found from another DFT+$U$ study of bulk LuNiO$_3$\cite{Zhuoran2015}. This issue will be further elaborated in the following sections. The different features found in between LNO/LAO and LNO/STO are discussed in Sec.~\ref{sec:STO}.

\section{LNO/LAO SL}
\label{sec:LAO}

As shown in Fig.~\ref{fig:PD}(a), the tensile-strained LNO/LAO shows
three distintive structural phases. The first one has no NiO$_6$
octahedral distortion except for the $(a^- a^- c^0)$-type tilting. This phase stabilizes in the weakly correlated regime of
$U_{\rm eff} \lesssim 1$ eV. The second is the CD phase found in the intermediately correlated ($1 \lesssim U_{\rm eff} \lesssim 3$ eV) and the less-strained ($a \lesssim 4$ \AA) region of phase diagram. Finally, the JT distorted phase is stable in the strongly correlated ($U_{\rm eff} \gtrsim 3$ eV) or strongly 
strained ($a \gtrsim 4$ \AA) regime. Hereafter we will denote the structures without any octahedral distortion by the `rotation-only' (RO) phase.

\begin{figure}[ht!]
  \centering
  \includegraphics[width=0.4\textwidth]{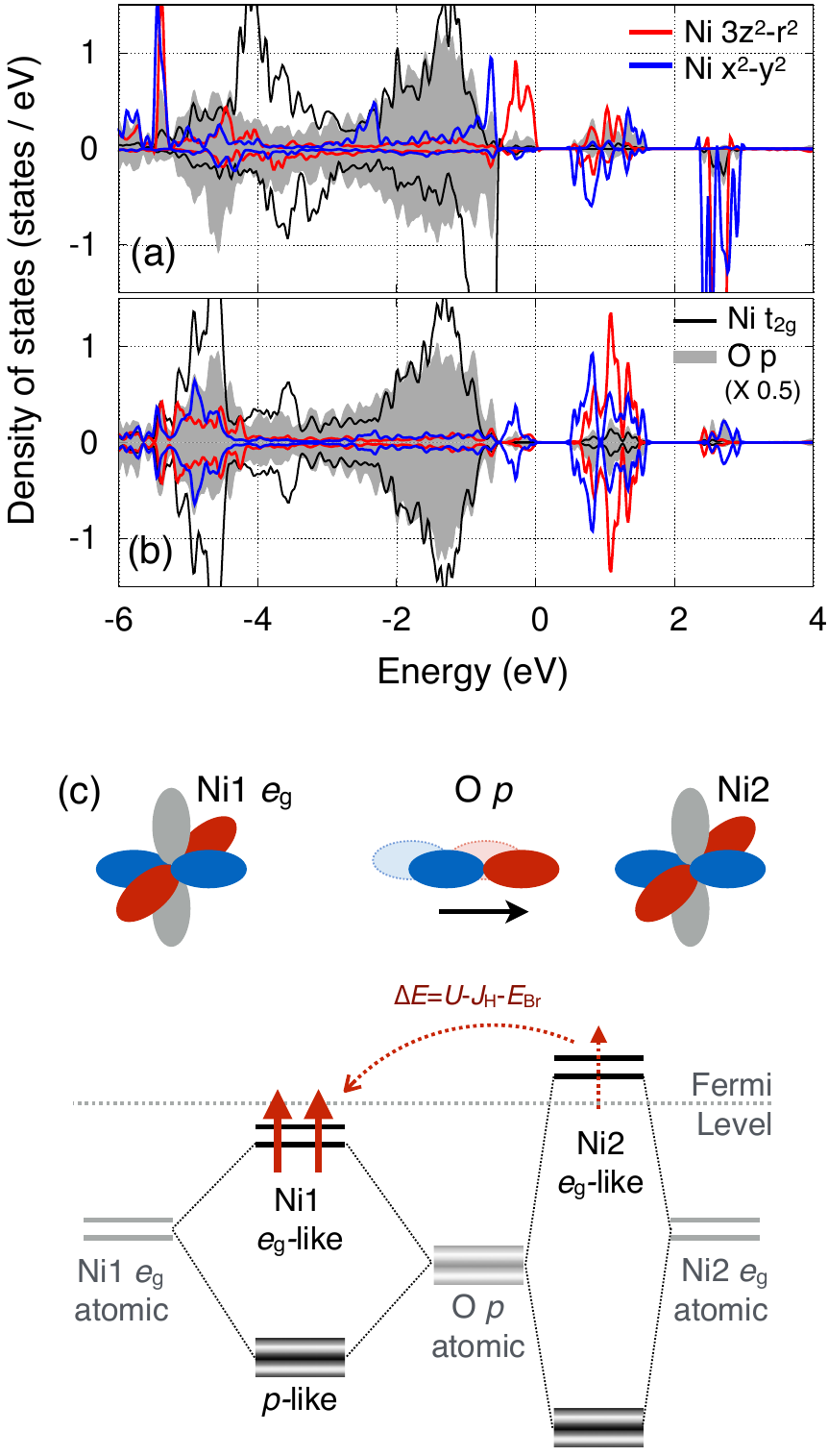}
  \caption{(Color online)
  (a,b) The calculated PDOS for Ni1 (a) and Ni2 (b) in LNO/LAO AF1+CD phase
  ($U_{\rm eff}$ = 3 eV and $a$ = 3.905\AA) (c) Schematic picture for stabilizing CD phase that involves the cooperation between Hund's coupling and breathing distortion (oxygen displacement as depicted in the figure). $E_{\rm Br}$
  denotes the energy gain from the distortion. 
  }
  \label{fig:LAO1}
\end{figure}

\subsubsection{CD phases with AF1, AF2, and FM orders}

Fig.~\ref{fig:LAO1}(a) and (b) present the calculated PDOS of AF1+CD phase
at a=3.905\AA~and $U_{\rm eff}$ = 3 eV. Band gap is clearly noticed.
Since other CD phases with different magnetic orders show the similar electronic structure, AF1+CD is chosen as the representative case. For the magnetic site of Ni1, the exchange splittings of $e_{\rm g}$ orbitals is well identified; see Fig.~\ref{fig:LAO1}(a). On the contrary, the non-magnetic Ni2 PDOS has the sizable portion of $e_{\rm g}$ peaks above the Fermi level as shown in Fig.~\ref{fig:LAO1}(b). Bader charge analysis \cite{Bader} yields about 0.09$e$ of the electron occupation difference between Ni1 and Ni2, which exhibits only small strain and $U_{\rm eff}$ dependence. This value is comparable with that in Ref.~\cite{Pentcheva2011PRB}, $\sim$0.07$e$. It is seen that oxygen $p$-orbital states are largely overlaped with Ni-$t_{\rm 2g}$ in both Ni1 and Ni2 sites.

Fig.~\ref{fig:LAO1}(c) shows the schematic picture of the underlying electronic cooperation for stabilizing the breathing-type  NiO$_6$ distortions. As discussed in
Ref.~\cite{Mazin2007PRL}, the CD costs the energy $\Delta E = U - J_{\rm H} - E_{\rm Br}$
where $E_{\rm Br}$ denotes the energy gain out of the breathing distortion. A recent DFT study shows that the breathing distortion is enhanced as Hund's coupling strength increases \cite{Park2015PRB}. It is also useful to recall here that the value of $U$ can likely 
be decreased by the Ni-$d$ hybridization with O-$p$-orbitals whereas $J_{\rm H}$ is known to be less sensitive to the delocalization of $e_{\rm g}$ electrons. Consequently, it becomes a plausible senario especially in the weakly-to-intermediately correlated regime that $\Delta E$ becomes negative and CD phase is energetically favored. In this situation, Ni2 would get more strongly hybridized with oxygen $p$ than Ni1. This picture can be further supported by the calculated planar Ni-O bond lengths; $d_{\rm planar}$(Ni1-O) and $d_{\rm planar}$(Ni2-O) is increased by 
7.8 and 2.5\%, respectively, as $a$ changes from 3.905 to 4.100~\AA. In the strongly correlated regime, on the other hand, CD phase becomes less unstable owing to the large $U$ energy cost and the JT phase stabilized  as shown  in Fig. \ref{fig:PD}(a) \cite{Mazin2007PRL}.

This modulation of Ni-O hybridization strength is closely related to the site-selective Mott  picture suggested for bulk nickelates
\cite{Park2012PRL,Bayo2013PRL} in which the $d$-$p$ charge transfer energy $\Delta_{\rm ct} \equiv \epsilon_d - \epsilon_p$ is negative and the magnetic inactivity of Ni2 $e_{\rm g}$ orbital 
is attributed to the strong hybridization with the oxygen $p$ ligand holes, 
$d^8\bar{L}^2$ \cite{Mizokawa2000PRB,Johnston2014PRL,Green2016}. It may be possible that our SL geometry belongs to a different case. Also, Ref.~\cite{Pascut2017} emphasizes the importance of the chosen energy range and argues that the $d^{7+\delta}$--$d^{7-\delta}$ CD picture is more plausible in the low-energy range ($\epsilon_F -3  \lesssim E \lesssim \epsilon_F + 3$ eV). Although our PDOS of Ni2-$e_{\rm g}$--O-$p$ hybrid states locating at +1eV is seemingly more consistent with  $d^{7+\delta}$--$d^{7-\delta}$ picture and with the low-energy model suggested in Ref.~\cite{Subedi2015}, we believe that, the static limitation of our current computation method prevents us from making the detailed comparisons regarding this multiplet physics.

Lastly, we briefly comment on AF2+CD and FM+CD phase. The latter was reported 
to be the ground state in the previous DFT+$U$ calculations \cite{Pentcheva2011PRB,Park2012PRL}. 
The electronic structure of AF2+CD is not much different from that of AF1+CD in spite of their total energy difference. In FM+CD phase, on the other hand, Ni2 shows a finite, albeit small, exchange splitting unlike the other AF phases.
%In addition, Bader charge and magnetic moment differences between Ni1 and Ni2 sites
%in AF1+CD and FM+CD phases show noticeably different behavior under tensile strain;
%while the Bader charge and magnetic moment
%differences slightly increase as tensile strain is enhanced in the AF1+CD phase ($\Delta e=0.092
%\rightarrow 0.095e$ and $\Delta \mu=1.30 \rightarrow 1.40\mu_{\rm B}$, respectively), 
%in FM+CD phase they show the opposite behavior ($\Delta e=0.087
%\rightarrow 0.043e$ and $\Delta \mu=0.74 \rightarrow 0.32\mu_{\rm B}$, respectively).
%Whether such behavior implies different nature of the CD in those two magnetic phases
%is unclear 
%Although such behavior implies the stronger CD in the AF1+CD phase, 
%the origin of the magnetism dependence remains unclear due to the fuzzy distinction between 
%the itinerant and the local picture in this system, which deserves further theoretical investigations. 

\subsubsection{JT phases with AF3, AF4, and FM orders}

\begin{figure}[ht!]
	\centering
	\includegraphics[width=0.4\textwidth]{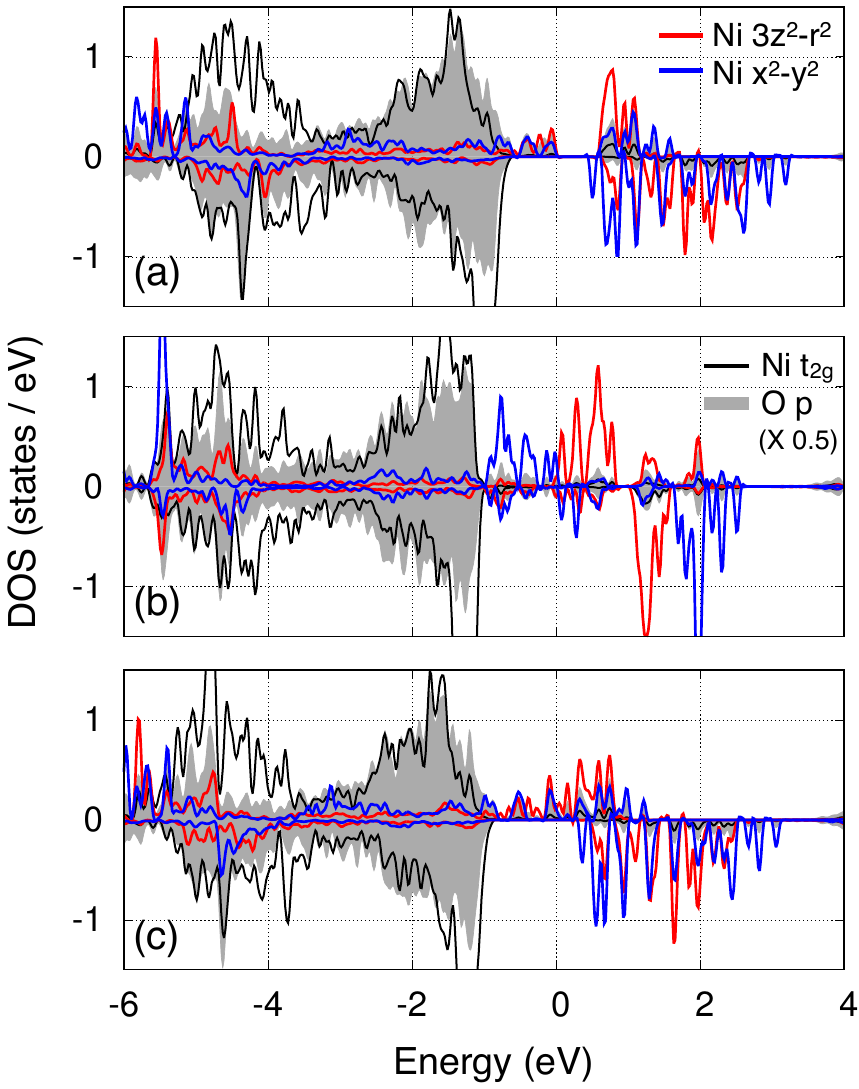}
	\caption{(Color online) The calculated PDOS for (a) FM+JT, (b) AF4, and (c) FM phase
		with $U_{\rm eff}$ = 3 eV and $a$ = 3.905\AA.
	}
	\label{fig:LAO2}
\end{figure}

In Fig.~\ref{fig:LAO2}(a), we present the PDOS of FM+JT phase calculated at a=3.905\AA~ and $U_{\rm eff}$=3 eV as the representative case of JT phases. Contrary to the CD case, 
the $d_{3z^2-r^2}$ and $d_{x^2-y^2}$ orbital characters are mixed with each other due 
to the formation of $d_{3x^2-r^2} / d_{3y^2-r^2}$ orbital order (note that here we used the global coordinate axes to define $d_{3x^2-r^2}$ and $d_{3y^2-r^2}$). The octahedral distortion reduces the overlap integral between $d_{3x^2-r^2}$ and $d_{3y^2-r^2}$, and further facilitates the minimization of 
Coulomb repulsion energy. Both Ni1 and Ni2 sites are magnetic, and the sizable $e_{\rm g}$-orbital
exchange splittings are clearly noticed. While FM+JT phase is the ground state in large-$U_{\rm eff}$ regime as shown in Fig. \ref{fig:PD}(a), AF3+JT phase is most stable when 
$U_{\rm eff} \simeq 2$ eV and a $>$ 4 \AA. The energy differences, however, are very small in this regime; FM+JT and AF4+JT phases have the higher energy than AF3+JT by about 2 to 3 meV per formula unit at $U_{\rm eff} = 2$ eV and a=4.1 \AA.

\subsubsection{RO phases with AF4 and FM orders}

Fig.~\ref{fig:LAO2}(b) and (c) shows the electronic structure of RO phase of AF4 and FM, respectively, calculated at a=3.905\AA~ and $U_{\rm eff}$ = 3 eV. Note that these RO phases are metastable having the higher energies than JT or CD phases. Both are metallic in this parameter set while AF4 becomes insulating by applying either tensile strain or imposing larger $U_{\rm eff}$. Well separated $e_{\rm g}$ states are apparent in Fig.~\ref{fig:LAO2}(b).

\begin{figure}[t]
  \centering
  \includegraphics[width=0.45\textwidth]{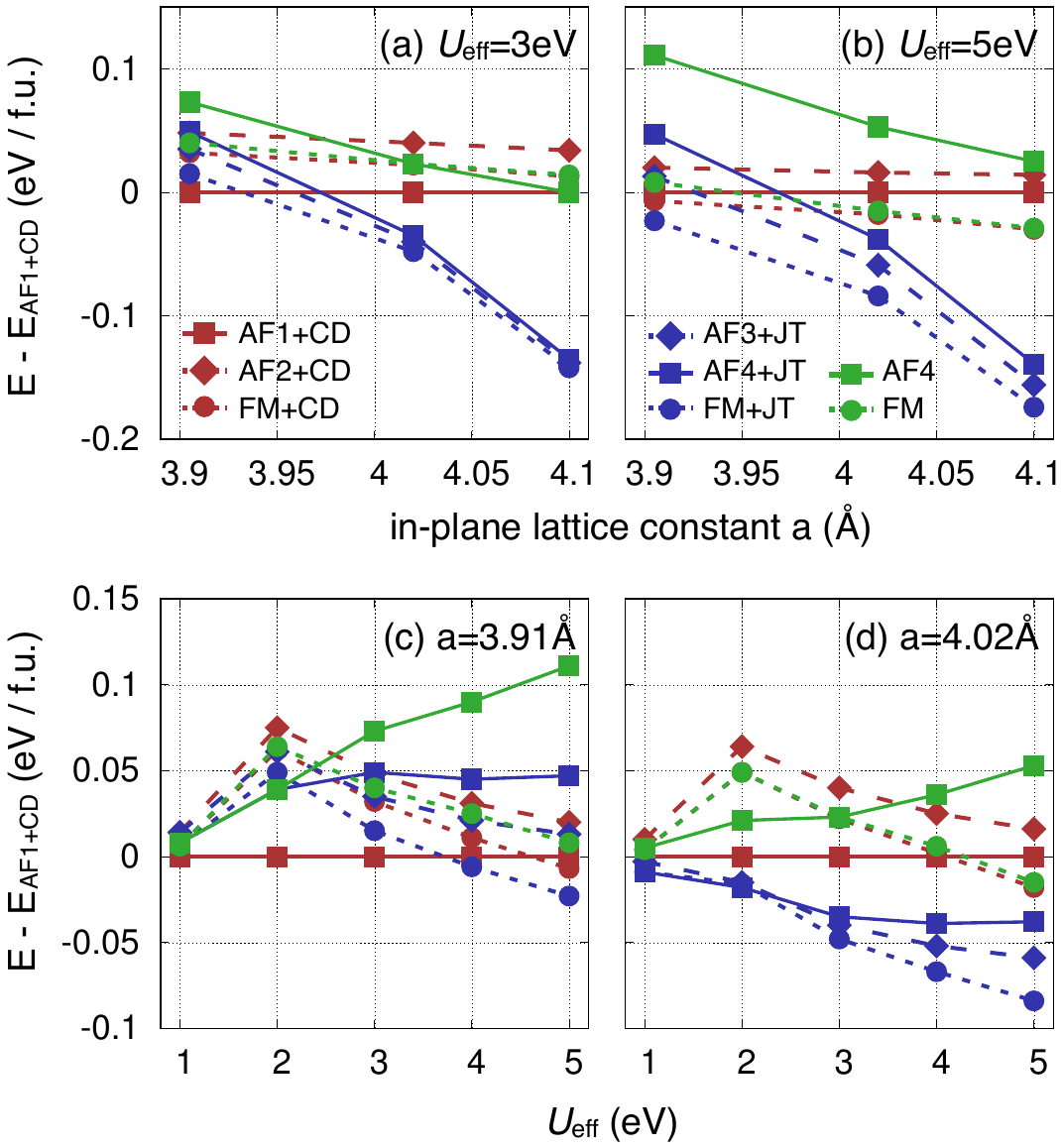}
  \caption{(Color online)
  Relative energy differences with respect to AF1+CD phase as a function 
  (a,b) strain and (c,d) $U_{\rm eff}$ value. Three different structure types of CD, JT and RO are depicted by red, blue and green, respectively.
  }
  \label{fig:LAO3}
\end{figure}

\subsubsection{Energy differences in the $U_{\rm eff}$-strain space}

Fig.~\ref{fig:LAO3}(a) summarizes the total energy calculation results by varying lattice constant with a fixed $U_{\rm eff}$=3 eV. It shows that JT structure is favored over CD phase in the presence of sizable tensile strain. It can be attributed to the reduced $d$-$p$ hybridization, resulting in the smaller energy gain $E_{\rm Br}$ (see Fig.~\ref{fig:LAO1}(c)) and facilitating the formation of JT phase with localized $d_{3x^2-r^2} / d_{3y^2-r^2}$ orbitals. At a=4.10 \AA, the calculated energy differences between JT and other phases are as large as $\sim$0.15 eV / f.u. which dominates over the magnetic energy scales. While AF4-RO phase gets
lower in energy as the larger tensile strain is introduced (green line), JT phases still remain as the lowest energy configuration. At the smallest value of strain ($a$=3.9), AF1+CD
is the ground state. Under compressive strains, it was reported that CD phase tends to be suppressed
and the system become metallic \cite{Pentcheva2011PRB}. The similar feature is found at a larger value of $U_{\rm eff}$=5 eV; see Fig.~\ref{fig:LAO3}(b). An interesting difference is that JT phases
are further stabilized and the FM+JT becomes the ground state even at a=3.905\AA. 
%It can be seen that, within the same type of structural distortion, FM orders become more lower in energy than the other magnetic orders. In addition, the FM (RO) phase tends to be favored over the CD phases when $U_{\rm eff}$ is enhanced, while AF phase become unfavorable.

In Fig.~\ref{fig:LAO3}(c) and (d), the calculated total energies are presented as a function of $U_{\rm eff}$ at the fixed values of strain. At a=3.905\AA, AF1+CD phase remains as the ground state
for $U_{\rm eff} \leq 3$ eV whereas FM+JT becomes more stable at $U_{\rm eff}\geq$ 4 eV (see Fig.~\ref{fig:LAO3}(c)). It is also interesting to note that the relative energies increase at $U_{\rm eff}\leq$2 eV, and start to mostly decrease at the larger $U_{\rm eff}$. Applying the greater tensile strains makes the JT phases energetically more favorable as clearly seen in Fig.~\ref{fig:LAO3}(d). It is consistent with the previous studies of two-orbital Kugel-Khomskii model \cite{Khaliullin2005PTPS,Chaloupka2008PRL}.

%Finally, we would like to comment on the inconsistency between our results to the previous 
%theoretical result\cite{Chaloupka2008PRL}, where the tensile strain is predicted to stabilize
%the $d_{x^2-y^2}$-orbital polarized phase with in-plane antiferromagnetic order. In our calculation,
%on the contrary, the energy difference between the FM+JT and the RO+AF phases at $U_{\rm eff}$ = 5 eV
%increases from 0.2 to 0.3 eV when $a$ is increased from 3.905 to 4.10 \AA. 
%With extremely large tensile strain it is possible that the $d_{x^2-y^2}$-orbital order becomes 
%the ground state, as shown in a LaTiO$_3$/LaAlO$_3$ SL where large tensile strain induces 
%transition from a JT order to a $d_{xy}$-ordered phase\cite{Alex2014PRB}. Also, there has been a
%work pointing out the possible overestimation of the Hund's coupling in DFT+$U$ 
%calculations\cite{Park2015PRB}. Treatment of the Hund's coupling in a different way 
%can affect the stability between different structural and orbital ordered states, which we leave as 
%an interesting future subject.

\section{LNO/STO SL and the substrate effect}
\label{sec:STO}

\begin{figure*}
  \centering
  \includegraphics[width=1.0\textwidth]{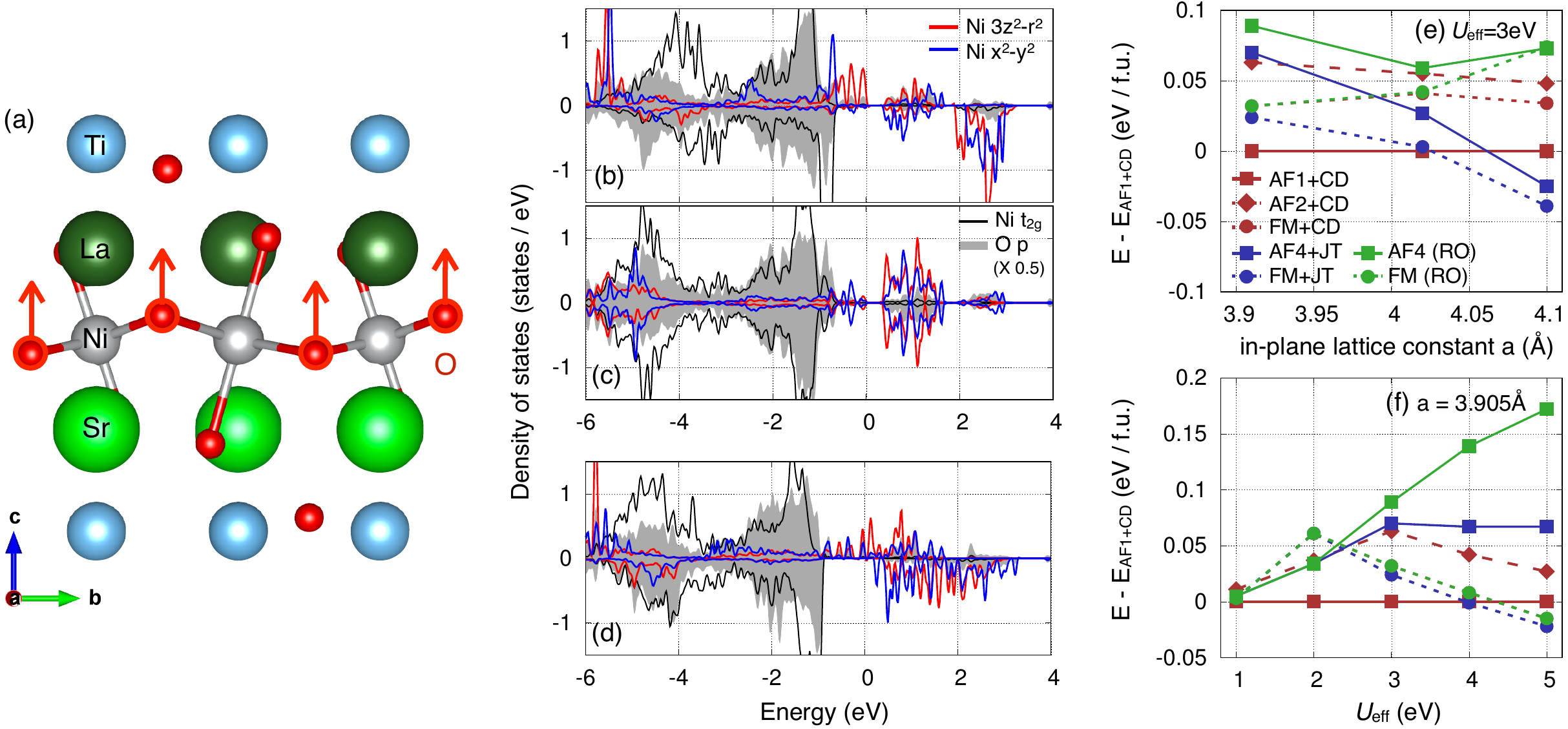}
  \caption{(Color online)
  (a) The optimized atomic structure of LNO/STO SL. The direction of oxygen (red spheres) displacements with respect to LNO/LAO is highlighted by red arrows.
  (b,c) Ni-$3d$ PDOS for Ni1 (b) and Ni2 (c) in AF1+CD phase calculated with
  $U_{\rm eff}$ = 3 eV and $a$ = 3.905\AA.  (d) PDOS for the FM+JT phase 
  when $U_{\rm eff}$ = 3 eV and $a$ = 3.905\AA.
  (e,f) The calculated total energy differences with respect to AF1+CD  
   by varying (e) $U_{\rm eff}$ and (f) strain. 
  }
  \label{fig:STO}
\end{figure*}

While the phase diagram of LNO/STO is qualitatively similar with LNO/LAO (see Fig.~\ref{fig:PD}), a notable difference is the stabilization of AF1+CD phase persisting up to the larger tensile strain regime. As discussed further below, this expansion of the AF1+CD area, or the shrinkage of JT, can be attributed to the internal electric field effect caused by the chemical charge imbalance between La$^{3+}$O and Sr$^{2+}$O layers. Note that the polarity-induced band bending reported in a previous study \cite{Geisler2017} is not relevant to our LNO$_1$/STO$_1$ SL setup.

Fig.~\ref{fig:STO}(b) and (c) presents the calculated PDOS for Ni1 and Ni2, respectively, which shows the CD between two sites. Bader charge analysis shows that the electron number difference is  comparable to that of LNO/LAO, and slightly increases up to 0.11$e$ as $a$ increases. Here we note that, contrary to the LNO/LAO SL, the nominal charge imbalance between (LaO)$^{1+}$ and (SrO)$^{0}$ layers can induces the electric field for (NiO$_2$)$^{1-}$ layer sandwiched in between. In response to this internal electric field, further structural relaxations can occur, and in fact, oxygen 
displacements along the $\hat{c}$-direction are observed as presented in Fig.~\ref{fig:STO}(a). Its effect can be noticed in PDOS; compared to LNO/LAO (see Fig.~\ref{fig:LAO1}(a) and (b)), the $d_{3z^2-r^2}$ state are lower in energy by $\sim$0.2 eV than $d_{x^2-y^2}$ in the unoccupied bands. Below the Fermi level, the $d_{3z^2-r^2}$ weight is mixed into the $d_{x^2-y^2}$ peak located around $-$0.8 eV while $d_{x^2-y^2}$-orbital is still lower in its energy than $d_{3z^2-r^2}$ presumably due to the larger tetragonal distortion at the Ni1 site.

Fig.~\ref{fig:STO}(d) shows the PDOS for the FM+JT phase where the vanishing size of gap is noted, which is in contrast to the case of LNO/LAO (Fig.~\ref{fig:LAO1}(c)). It implies that the energy gain by gap opening is smaller in LNO/STO than in LNO/LAO SL. Note that JT distortion induces the $d_{3x^2-r^2}$/$d_{3y^2-r^2}$ orbital order, and the internal-field-induced $e_{\rm g}$ splitting can suppress the JT distortion. This interpretation is consistent with the result of strain dependence. The oxygen displacement along $\hat{c}$-direction increases from 0.06 to 0.11 \AA~ as $a$ increases from 3.905 to 4.100 \AA, enhancing the $e_{\rm g}$ splitting.

Fig.~\ref{fig:STO}(e) summarizes the relative energy differences of the spin and lattice configurations considered in this study. The relative stabilization energy of JT phases with respect to AF1+CD is substantially increased as $a$ increases. See also Fig.~\ref{fig:LAO3}(a) for comparison. AF1+CD is favored when $a \leq$ 4.020 \AA. The $U_{\rm eff}$ dependence is also investigated whose results are presented in Fig.~\ref{fig:STO}(f). Almost the same behavior is observed in LNO/LAO (see Fig.~\ref{fig:LAO3}(c)).

\section{Orbital polarizations in LNO/LAO and LNO/STO SL}
\label{sed:OP}

\begin{figure*}
  \centering
  \includegraphics[width=0.98\textwidth]{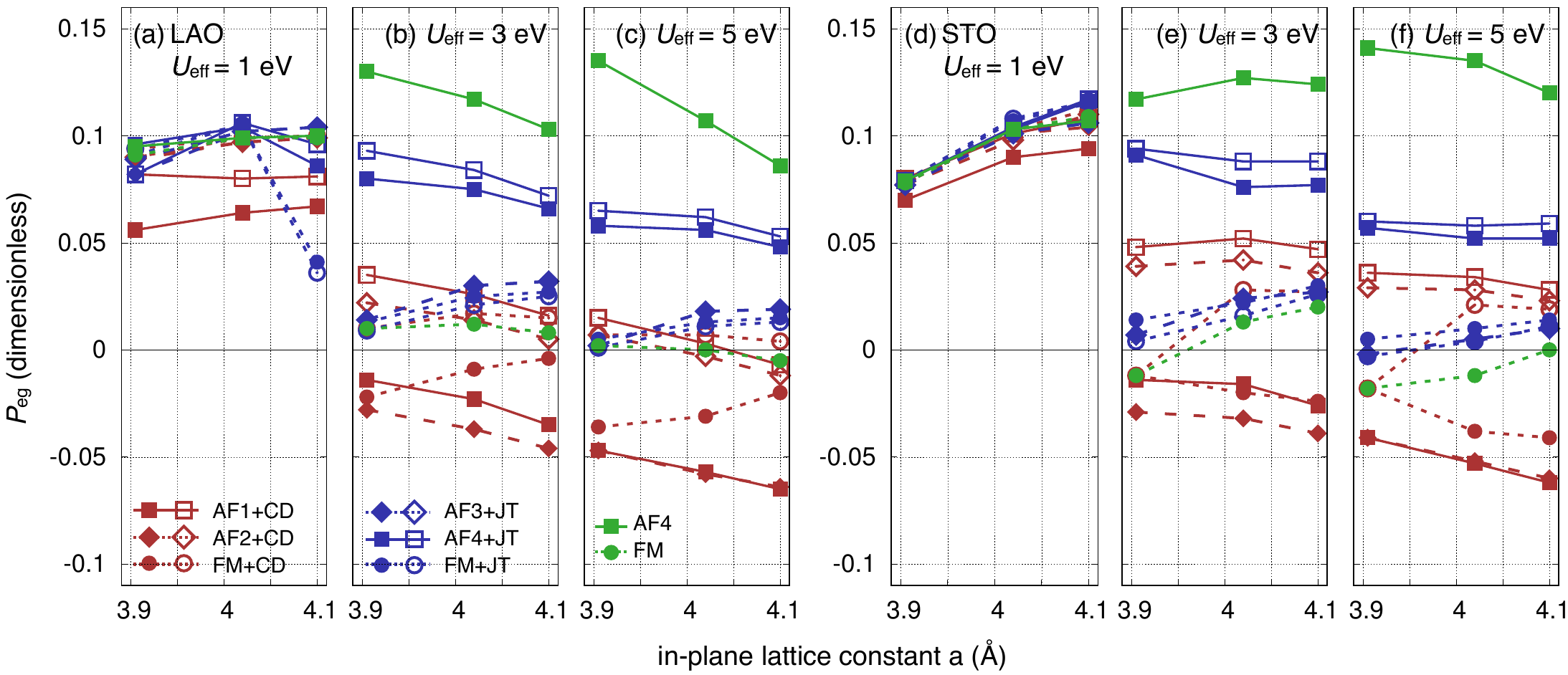}
  \caption{(Color online)
  The calculated orbital polarizations of LNO/LAO (a, b, c) and LNO/STO (d, e, f) in different magnetic and structural configurations as a function of in-plane lattice constant. Three different $U_{\rm eff}$ values are taken into account; (a,d) 1 eV, (b,e) 3 eV and (c,f) 5 eV.
  The filled and empty symbols depict the results of Ni1 and Ni2, respectively, in CD and JT. Note that, at $U_{\rm eff}$ = 1 eV, all the systems are metallic whereas they are all gapped at $U_{\rm eff}$ = 3 and 5 eV but the FM phase.
  }
  \label{fig:OP}
\end{figure*}

Orbital polarization has been a key quantity in the nickelate study which can be defined as 
\begin{align*}
P_{e_{\rm g}} &= \frac{n_{d_{x^2-y^2}} - n_{d_{3z^2-r^2}}}{n_{d_{x^2-y^2}} + n_{d_{3z^2-r^2}}},
\end{align*}
where $n_{d_{x^2-y^2}}$ and $n_{d_{3z^2-r^2}}$ refers to  the electron occupation in $d_{x^2-y^2}$ and $d_{3z^2-r^2}$ orbital, respectively. It can be also defined in terms of hole densities as measured by XAS \cite{Wu2013,Peil2014PRB,Park2016PRB}. From the crystal-field point of view, tensile strain is expected to lower the $d_{x^2-y^2}$ on-site energy and to enhance $P_{e_{\rm g}}$ as reported in Ref.~\cite{Wu2013}. It is supported by two recent DFT+DMFT studies on a tensile-strained film\cite{Peil2014PRB} and a LNO$_4$/LAO$_4$ superlattice\cite{Peil2014PRB,Park2016PRB}, both of which incorporate the octahedral rotations and tiltings but not the magnetism or CD/JT distortions. It was shown that octahedral tilting under tensile strain enhances $P_{e_{\rm g}}$ when the system is paramagnetic and metallic, and that  on-site correlations reduce $P_{e_{\rm g}}$ \cite{Peil2014PRB}. It is also consistent with a previous DMFT study \cite{MJH2011PRL}. On the other hand, another X-ray linear dichroism experiment of LNO thin films on top of LAO and LNO$_1$/LAO$_1$ SL reports no enhancement of 
$P_{e_{\rm g}}$ under tensile strain \cite{Chakhalian:2011ef,Freeland2011}, possibly due to the
octahedral breathing modes introducing CD \cite{Chakhalian:2011ef,Pentcheva2011PRB}. 
Another recent study on PrNiO$_3$/PrAlO$_3$ (whose bulk phase is known to be 
paramagnetic and metallic just like bulk LNO) reports the CD formation and the concomitant reduction of $P_{e_{\rm g}}$ under tensile strain \cite{Wu2015PRB}. Not achieved yet is the clear understanding of the $P_{e_{\rm g}}$ behavior in response to various control parameters.

To investigate its dependence on $U_{\rm eff}$, strain, and structural/magnetic configurations, we performed the extensive calculations whose results are summarized in Fig.~\ref{fig:OP}. At $U_{\rm eff}$ = 1 eV where systems remain metallic (Fig.~\ref{fig:OP}(a) and (d)), $P_{e_{\rm g}}$ is positive and mostly $\sim$0.1. It tends to increase as the greater tensile strain is applied. The exception is the LNO/LAO JT phases at a = 4.1\AA~ in which JT distortion becomes dominating. This result is therefore qualitatively consistent with Ref.~\cite{Wu2013,Peil2014PRB,Park2016PRB}. At larger $U_{\rm eff}$, $P_{e_{\rm g}}$ is suppressed significantly by CD or JT distortion which is also partly observed in a previous DFT+$U$ study of incorporating CD \cite{Pentcheva2011PRB}. The distinction between Ni1 and NI2 sites in CD phases is clearly seen in Fig.~\ref{fig:OP}(b,c,e,f), where AF1+CD and AF2+CD phases show the similar behavior of decreasing $P_{e_{\rm g}}$. In the AF4+JT phase, $P_{e_{\rm g}}$ remains to be 0.05--0.1 for $3 < U_{\rm eff} < 5$ eV, which shows the coupling between the $d_{x^2-y^2}$-type orbital ordering and the $\sqrt{2} \times \sqrt{2}$ checkerboard-type AF order. Similar conclusion can be made by comparing AF4 with FM phases (having neither CD nor JT order) \cite{Han:2012bc}. As mentioned above, AF4-type order favors the positive $P_{e_{\rm g}}$ in Ni $d^7$ configuration, even in the presence of JT order, while other configurations with FM nearest-neighbors show the vanishing $P_{e_{\rm g}}$. Contrary to the naive expectation based on crystal field, $P_{e_{\rm g}}$ tends to decrease as a function of tensile strain especially when $U_{\rm eff} \geq 3$ eV. It is attributed to the tetragonal splitting between $d_{x^2-y^2}$ and $d_{3z^2-r^2}$ being compensated by octahedral tiltings as previously discussed in Ref.~\cite{Peil2014PRB}. While there is no signifiacnt qualitative difference observed in between LAO (Fig.~\ref{fig:OP}(a-c)) and STO (Fig. \ref{fig:OP} (d-f)), the positive $P_{e_{\rm g}}$ is more favored in LNO/STO than in LNO/LAO for all $U_{\rm eff}$ values. It is possibly due to the combination of octahedral tiltings and oxygen displacements shown in Fig.~\ref{fig:STO}(a).

\section{Discussion and Summary}
\label{sed:discussion}

While our DFT$+U$ results provide the useful information, care needs to be taken especially when directly compared with the results obtained from experiments and other computation methods  \cite{Wu2013,Peil2014PRB,Park2016PRB,Chakhalian:2011ef,Freeland2011,Wu2015PRB}. There have been a couple of theoretical suggestions based on spin-density wave (SDW) picture that seems to reasonably well describe some aspects of insulating and magnetic states of SL \cite{SBLee2011PRL,SBLee2011PRB}. In Ref.~\cite{Frano2013PRL}, for example, an insulating phase carrying a non-collinear magnetic order was reported in two-unit-cell-thick SL being consistent with the SDW scenario. DFT+DMFT\cite{MJH2011PRL} and quasiparticle self-consistent GW calculations \cite{Han:2014be} on LNO/LAO also reports the strong $p$-orbital character persisting in the Fermi surface. Recent ARPES measurements indeed reported the well-observed Fermi surfaces in the thin-film LNO down to two unitcell\cite{King2014NNT,Yoo2016PRB}. Importantly, the significant effects of correlations can also be manifested as the band broadening and the disappearance of Fermi surface in the thinner limit \cite{King2014NNT}.

%DFT+$U$ vs. SDFT+$U$
%ARPES exp. by HK Yoo
%Limitation of the current work, limitation of DFT+$U$ in nickelate

In conclusion, we studied the ground state properties of tensile-strained LNO$_1$/LAO$_1$ and LNO$_1$/STO$_1$ SLs. Within DFT+$U$ method, we systematically investigated the wide region of phase space including the previously unexplored. In the presence of moderate Coulomb interaction
($U_{\rm eff} \sim 3$ eV) and tensile strain (a $\lesssim$ 4 \AA), an AF spin order with 
$2\sqrt{2} \times 2\sqrt{2}$ periodicity is found to be stabilized together with CD configuration. The stronger tensile strain and $U_{\rm eff}$ favor the JT-ordered phases. Our findings hopefully provide the useful information to understand ultrathin nickelates, not only the intriguing experimental data but also the seemingly different theoretical/computational results.

\section{Acknowledgements}
This work was supported by the National Research Foundation of Korea (NRF) grant funded by the Korea government (MSIT) (Grant No. NRF-2021R1A2C1009303, NRF-2018M3D1A1058754, and NRF-2020R1C1C1005900).

\bibliography{LNOLAOSTO}{}

\begin{thebibliography}{65}%
\makeatletter
\providecommand \@ifxundefined [1]{%
 \@ifx{#1\undefined}
}%
\providecommand \@ifnum [1]{%
 \ifnum #1\expandafter \@firstoftwo
 \else \expandafter \@secondoftwo
 \fi
}%
\providecommand \@ifx [1]{%
 \ifx #1\expandafter \@firstoftwo
 \else \expandafter \@secondoftwo
 \fi
}%
\providecommand \natexlab [1]{#1}%
\providecommand \enquote  [1]{``#1''}%
\providecommand \bibnamefont  [1]{#1}%
\providecommand \bibfnamefont [1]{#1}%
\providecommand \citenamefont [1]{#1}%
\providecommand \href@noop [0]{\@secondoftwo}%
\providecommand \href [0]{\begingroup \@sanitize@url \@href}%
\providecommand \@href[1]{\@@startlink{#1}\@@href}%
\providecommand \@@href[1]{\endgroup#1\@@endlink}%
\providecommand \@sanitize@url [0]{\catcode `\\12\catcode `\$12\catcode
  `\&12\catcode `\#12\catcode `\^12\catcode `\_12\catcode `\%12\relax}%
\providecommand \@@startlink[1]{}%
\providecommand \@@endlink[0]{}%
\providecommand \url  [0]{\begingroup\@sanitize@url \@url }%
\providecommand \@url [1]{\endgroup\@href {#1}{\urlprefix }}%
\providecommand \urlprefix  [0]{URL }%
\providecommand \Eprint [0]{\href }%
\providecommand \doibase [0]{https://doi.org/}%
\providecommand \selectlanguage [0]{\@gobble}%
\providecommand \bibinfo  [0]{\@secondoftwo}%
\providecommand \bibfield  [0]{\@secondoftwo}%
\providecommand \translation [1]{[#1]}%
\providecommand \BibitemOpen [0]{}%
\providecommand \bibitemStop [0]{}%
\providecommand \bibitemNoStop [0]{.\EOS\space}%
\providecommand \EOS [0]{\spacefactor3000\relax}%
\providecommand \BibitemShut  [1]{\csname bibitem#1\endcsname}%
\let\auto@bib@innerbib\@empty
%</preamble>
\bibitem [{\citenamefont {Torrance}\ \emph {et~al.}(1992)\citenamefont
  {Torrance}, \citenamefont {Lacorre}, \citenamefont {Nazzal}, \citenamefont
  {Ansaldo},\ and\ \citenamefont {Niedermayer}}]{RNOPD}%
  \BibitemOpen
  \bibfield  {author} {\bibinfo {author} {\bibfnamefont {J.~B.}\ \bibnamefont
  {Torrance}}, \bibinfo {author} {\bibfnamefont {P.}~\bibnamefont {Lacorre}},
  \bibinfo {author} {\bibfnamefont {A.~I.}\ \bibnamefont {Nazzal}}, \bibinfo
  {author} {\bibfnamefont {E.~J.}\ \bibnamefont {Ansaldo}},\ and\ \bibinfo
  {author} {\bibfnamefont {C.}~\bibnamefont {Niedermayer}},\ }\bibfield
  {title} {\bibinfo {title} {Systematic study of insulator-metal transitions in
  perovskites \textit{R} ${\mathrm{nio}}_{3}$ ( \textit{R} =pr,nd,sm,eu) due to
  closing of charge-transfer gap},\ }\href
  {https://doi.org/10.1103/PhysRevB.45.8209} {\bibfield  {journal} {\bibinfo
  {journal} {Phys. Rev. B}\ }\textbf {\bibinfo {volume} {45}},\ \bibinfo
  {pages} {8209} (\bibinfo {year} {1992})}\BibitemShut {NoStop}%
\bibitem [{\citenamefont {Medarde}(1997)}]{RNOreview}%
  \BibitemOpen
  \bibfield  {author} {\bibinfo {author} {\bibfnamefont {M.~L.}\ \bibnamefont
  {Medarde}},\ }\bibfield  {title} {\bibinfo {title} {Structural, magnetic and
  electronic properties of $r$nio$_3$ perovskites ($r$ = rare earth)},\ }\href
  {http://stacks.iop.org/0953-8984/9/i=8/a=003} {\bibfield  {journal} {\bibinfo
   {journal} {Journal of Physics: Condensed Matter}\ }\textbf {\bibinfo
  {volume} {9}},\ \bibinfo {pages} {1679} (\bibinfo {year} {1997})}\BibitemShut
  {NoStop}%
\bibitem [{\citenamefont {Middey}\ \emph {et~al.}(2016)\citenamefont {Middey},
  \citenamefont {Chakhalian}, \citenamefont {Mahadevan}, \citenamefont
  {Freeland}, \citenamefont {Millis},\ and\ \citenamefont
  {Sarma}}]{RNOh_review}%
  \BibitemOpen
  \bibfield  {author} {\bibinfo {author} {\bibfnamefont {S.}~\bibnamefont
  {Middey}}, \bibinfo {author} {\bibfnamefont {J.}~\bibnamefont {Chakhalian}},
  \bibinfo {author} {\bibfnamefont {P.}~\bibnamefont {Mahadevan}}, \bibinfo
  {author} {\bibfnamefont {J.}~\bibnamefont {Freeland}}, \bibinfo {author}
  {\bibfnamefont {A.}~\bibnamefont {Millis}},\ and\ \bibinfo {author}
  {\bibfnamefont {D.}~\bibnamefont {Sarma}},\ }\bibfield  {title} {\bibinfo
  {title} {Physics of ultrathin films and heterostructures of rare-earth
  nickelates},\ }\href {https://doi.org/10.1146/annurev-matsci-070115-032057}
  {\bibfield  {journal} {\bibinfo  {journal} {Annual Review of Materials
  Research}\ }\textbf {\bibinfo {volume} {46}},\ \bibinfo {pages} {305}
  (\bibinfo {year} {2016})}\BibitemShut {NoStop}%
\bibitem [{\citenamefont {Catalano}\ \emph {et~al.}(2018)\citenamefont
  {Catalano}, \citenamefont {Gibert}, \citenamefont {Fowlie}, \citenamefont
  {Íñiguez}, \citenamefont {Triscone},\ and\ \citenamefont
  {Kreisel}}]{Catalano2018_Review}%
  \BibitemOpen
  \bibfield  {author} {\bibinfo {author} {\bibfnamefont {S.}~\bibnamefont
  {Catalano}}, \bibinfo {author} {\bibfnamefont {M.}~\bibnamefont {Gibert}},
  \bibinfo {author} {\bibfnamefont {J.}~\bibnamefont {Fowlie}}, \bibinfo
  {author} {\bibfnamefont {J.}~\bibnamefont {Íñiguez}}, \bibinfo {author}
  {\bibfnamefont {J.-M.}\ \bibnamefont {Triscone}},\ and\ \bibinfo {author}
  {\bibfnamefont {J.}~\bibnamefont {Kreisel}},\ }\bibfield  {title} {\bibinfo
  {title} {{Rare-earth nickelates RNiO3: thin films and heterostructures}},\
  }\href {https://doi.org/10.1088/1361-6633/aaa37a} {\bibfield  {journal}
  {\bibinfo  {journal} {Reports on Progress in Physics}\ }\textbf {\bibinfo
  {volume} {81}},\ \bibinfo {pages} {046501} (\bibinfo {year}
  {2018})}\BibitemShut {NoStop}%
\bibitem [{\citenamefont {Li}\ \emph {et~al.}(2019)\citenamefont {Li},
  \citenamefont {Lee}, \citenamefont {Wang}, \citenamefont {Osada},
  \citenamefont {Crossley}, \citenamefont {Lee}, \citenamefont {Cui},
  \citenamefont {Hikita},\ and\ \citenamefont {Hwang}}]{Li2019Nature}%
  \BibitemOpen
  \bibfield  {author} {\bibinfo {author} {\bibfnamefont {D.}~\bibnamefont
  {Li}}, \bibinfo {author} {\bibfnamefont {K.}~\bibnamefont {Lee}}, \bibinfo
  {author} {\bibfnamefont {B.~Y.}\ \bibnamefont {Wang}}, \bibinfo {author}
  {\bibfnamefont {M.}~\bibnamefont {Osada}}, \bibinfo {author} {\bibfnamefont
  {S.}~\bibnamefont {Crossley}}, \bibinfo {author} {\bibfnamefont {H.~R.}\
  \bibnamefont {Lee}}, \bibinfo {author} {\bibfnamefont {Y.}~\bibnamefont
  {Cui}}, \bibinfo {author} {\bibfnamefont {Y.}~\bibnamefont {Hikita}},\ and\
  \bibinfo {author} {\bibfnamefont {H.~Y.}\ \bibnamefont {Hwang}},\ }\bibfield
  {title} {\bibinfo {title} {Superconductivity in an infinite-layer
  nickelate},\ }\href {https://doi.org/10.1038/s41586-019-1496-5} {\bibfield
  {journal} {\bibinfo  {journal} {Nature}\ }\textbf {\bibinfo {volume} {572}},\
  \bibinfo {pages} {624} (\bibinfo {year} {2019})}\BibitemShut {NoStop}%
\bibitem [{\citenamefont {Nomura}\ and\ \citenamefont
  {Arita}(2022)}]{Nomura2022rev}%
  \BibitemOpen
  \bibfield  {author} {\bibinfo {author} {\bibfnamefont {Y.}~\bibnamefont
  {Nomura}}\ and\ \bibinfo {author} {\bibfnamefont {R.}~\bibnamefont {Arita}},\
  }\bibfield  {title} {\bibinfo {title} {Superconductivity in infinite-layer
  nickelates},\ }\href {https://doi.org/10.1088/1361-6633/ac5a60} {\bibfield
  {journal} {\bibinfo  {journal} {Reports on Progress in Physics}\ }\textbf
  {\bibinfo {volume} {85}},\ \bibinfo {pages} {052501} (\bibinfo {year}
  {2022})}\BibitemShut {NoStop}%
\bibitem [{\citenamefont {Sakakibara}\ \emph {et~al.}(2020)\citenamefont
  {Sakakibara}, \citenamefont {Usui}, \citenamefont {Suzuki}, \citenamefont
  {Kotani}, \citenamefont {Aoki},\ and\ \citenamefont
  {Kuroki}}]{Sakakibara2020}%
  \BibitemOpen
  \bibfield  {author} {\bibinfo {author} {\bibfnamefont {H.}~\bibnamefont
  {Sakakibara}}, \bibinfo {author} {\bibfnamefont {H.}~\bibnamefont {Usui}},
  \bibinfo {author} {\bibfnamefont {K.}~\bibnamefont {Suzuki}}, \bibinfo
  {author} {\bibfnamefont {T.}~\bibnamefont {Kotani}}, \bibinfo {author}
  {\bibfnamefont {H.}~\bibnamefont {Aoki}},\ and\ \bibinfo {author}
  {\bibfnamefont {K.}~\bibnamefont {Kuroki}},\ }\bibfield  {title} {\bibinfo
  {title} {Model construction and a possibility of cupratelike pairing in a new
  ${d}^{9}$ nickelate superconductor
  $(\mathrm{Nd},\mathrm{Sr}){\mathrm{nio}}_{2}$},\ }\href
  {https://doi.org/10.1103/PhysRevLett.125.077003} {\bibfield  {journal}
  {\bibinfo  {journal} {Phys. Rev. Lett.}\ }\textbf {\bibinfo {volume} {125}},\
  \bibinfo {pages} {077003} (\bibinfo {year} {2020})}\BibitemShut {NoStop}%
\bibitem [{\citenamefont {Jiang}\ \emph {et~al.}(2019)\citenamefont {Jiang},
  \citenamefont {Si}, \citenamefont {Liao},\ and\ \citenamefont
  {Zhong}}]{Jiang2019}%
  \BibitemOpen
  \bibfield  {author} {\bibinfo {author} {\bibfnamefont {P.}~\bibnamefont
  {Jiang}}, \bibinfo {author} {\bibfnamefont {L.}~\bibnamefont {Si}}, \bibinfo
  {author} {\bibfnamefont {Z.}~\bibnamefont {Liao}},\ and\ \bibinfo {author}
  {\bibfnamefont {Z.}~\bibnamefont {Zhong}},\ }\bibfield  {title} {\bibinfo
  {title} {Electronic structure of rare-earth infinite-layer
  $r\mathrm{Ni}{\mathrm{o}}_{2}(r=\mathrm{La},\mathrm{Nd})$},\ }\href
  {https://doi.org/10.1103/PhysRevB.100.201106} {\bibfield  {journal} {\bibinfo
   {journal} {Phys. Rev. B}\ }\textbf {\bibinfo {volume} {100}},\ \bibinfo
  {pages} {201106} (\bibinfo {year} {2019})}\BibitemShut {NoStop}%
\bibitem [{\citenamefont {Nomura}\ \emph {et~al.}(2019)\citenamefont {Nomura},
  \citenamefont {Hirayama}, \citenamefont {Tadano}, \citenamefont {Yoshimoto},
  \citenamefont {Nakamura},\ and\ \citenamefont {Arita}}]{Nomura2019}%
  \BibitemOpen
  \bibfield  {author} {\bibinfo {author} {\bibfnamefont {Y.}~\bibnamefont
  {Nomura}}, \bibinfo {author} {\bibfnamefont {M.}~\bibnamefont {Hirayama}},
  \bibinfo {author} {\bibfnamefont {T.}~\bibnamefont {Tadano}}, \bibinfo
  {author} {\bibfnamefont {Y.}~\bibnamefont {Yoshimoto}}, \bibinfo {author}
  {\bibfnamefont {K.}~\bibnamefont {Nakamura}},\ and\ \bibinfo {author}
  {\bibfnamefont {R.}~\bibnamefont {Arita}},\ }\bibfield  {title} {\bibinfo
  {title} {Formation of a two-dimensional single-component correlated electron
  system and band engineering in the nickelate superconductor
  ${\mathrm{ndnio}}_{2}$},\ }\href
  {https://doi.org/10.1103/PhysRevB.100.205138} {\bibfield  {journal} {\bibinfo
   {journal} {Phys. Rev. B}\ }\textbf {\bibinfo {volume} {100}},\ \bibinfo
  {pages} {205138} (\bibinfo {year} {2019})}\BibitemShut {NoStop}%
\bibitem [{\citenamefont {Ryee}\ \emph {et~al.}(2020)\citenamefont {Ryee},
  \citenamefont {Yoon}, \citenamefont {Kim}, \citenamefont {Jeong},\ and\
  \citenamefont {Han}}]{Ryee2020}%
  \BibitemOpen
  \bibfield  {author} {\bibinfo {author} {\bibfnamefont {S.}~\bibnamefont
  {Ryee}}, \bibinfo {author} {\bibfnamefont {H.}~\bibnamefont {Yoon}}, \bibinfo
  {author} {\bibfnamefont {T.~J.}\ \bibnamefont {Kim}}, \bibinfo {author}
  {\bibfnamefont {M.~Y.}\ \bibnamefont {Jeong}},\ and\ \bibinfo {author}
  {\bibfnamefont {M.~J.}\ \bibnamefont {Han}},\ }\bibfield  {title} {\bibinfo
  {title} {Induced magnetic two-dimensionality by hole doping in the
  superconducting infinite-layer nickelate
  ${\mathrm{nd}}_{1\ensuremath{-}x}{\mathrm{sr}}_{x}{\mathrm{nio}}_{2}$},\
  }\href {https://doi.org/10.1103/PhysRevB.101.064513} {\bibfield  {journal}
  {\bibinfo  {journal} {Phys. Rev. B}\ }\textbf {\bibinfo {volume} {101}},\
  \bibinfo {pages} {064513} (\bibinfo {year} {2020})}\BibitemShut {NoStop}%
\bibitem [{\citenamefont {Hansmann}\ \emph {et~al.}(2009)\citenamefont
  {Hansmann}, \citenamefont {Yang}, \citenamefont {Toschi}, \citenamefont
  {Khaliullin}, \citenamefont {Andersen},\ and\ \citenamefont
  {Held}}]{Hansmann2009PRL}%
  \BibitemOpen
  \bibfield  {author} {\bibinfo {author} {\bibfnamefont {P.}~\bibnamefont
  {Hansmann}}, \bibinfo {author} {\bibfnamefont {X.}~\bibnamefont {Yang}},
  \bibinfo {author} {\bibfnamefont {A.}~\bibnamefont {Toschi}}, \bibinfo
  {author} {\bibfnamefont {G.}~\bibnamefont {Khaliullin}}, \bibinfo {author}
  {\bibfnamefont {O.~K.}\ \bibnamefont {Andersen}},\ and\ \bibinfo {author}
  {\bibfnamefont {K.}~\bibnamefont {Held}},\ }\bibfield  {title} {\bibinfo
  {title} {Turning a nickelate fermi surface into a cupratelike one through
  heterostructuring},\ }\href {https://doi.org/10.1103/PhysRevLett.103.016401}
  {\bibfield  {journal} {\bibinfo  {journal} {Phys. Rev. Lett.}\ }\textbf
  {\bibinfo {volume} {103}},\ \bibinfo {pages} {016401} (\bibinfo {year}
  {2009})}\BibitemShut {NoStop}%
\bibitem [{\citenamefont {Boris}\ \emph {et~al.}(2011)\citenamefont {Boris},
  \citenamefont {Matiks}, \citenamefont {Benckiser}, \citenamefont {Frano},
  \citenamefont {Popovich}, \citenamefont {Hinkov}, \citenamefont {Wochner},
  \citenamefont {Castro-Colin}, \citenamefont {Detemple},\ and\ \citenamefont
  {Malik}}]{Boris:2011vt}%
  \BibitemOpen
  \bibfield  {author} {\bibinfo {author} {\bibfnamefont {A.~V.}\ \bibnamefont
  {Boris}}, \bibinfo {author} {\bibfnamefont {Y.}~\bibnamefont {Matiks}},
  \bibinfo {author} {\bibfnamefont {E.}~\bibnamefont {Benckiser}}, \bibinfo
  {author} {\bibfnamefont {A.}~\bibnamefont {Frano}}, \bibinfo {author}
  {\bibfnamefont {P.}~\bibnamefont {Popovich}}, \bibinfo {author}
  {\bibfnamefont {V.}~\bibnamefont {Hinkov}}, \bibinfo {author} {\bibfnamefont
  {P.}~\bibnamefont {Wochner}}, \bibinfo {author} {\bibfnamefont
  {M.}~\bibnamefont {Castro-Colin}}, \bibinfo {author} {\bibfnamefont
  {E.}~\bibnamefont {Detemple}},\ and\ \bibinfo {author} {\bibfnamefont
  {V.~K.}\ \bibnamefont {Malik}},\ }\bibfield  {title} {\bibinfo {title}
  {{Dimensionality control of electronic phase transitions in nickel-oxide
  superlattices}},\ }\href {https://doi.org/10.1126/science.1202647} {\bibfield
   {journal} {\bibinfo  {journal} {Science}\ }\textbf {\bibinfo {volume}
  {332}},\ \bibinfo {pages} {937} (\bibinfo {year} {2011})}\BibitemShut
  {NoStop}%
\bibitem [{\citenamefont {Gray}\ \emph {et~al.}(2011)\citenamefont {Gray},
  \citenamefont {Janotti}, \citenamefont {Son}, \citenamefont {LeBeau},
  \citenamefont {Ueda}, \citenamefont {Yamashita}, \citenamefont {Kobayashi},
  \citenamefont {Kaiser}, \citenamefont {Sutarto}, \citenamefont {Wadati},
  \citenamefont {Sawatzky}, \citenamefont {Van~de Walle}, \citenamefont
  {Stemmer},\ and\ \citenamefont {Fadley}}]{Gray2011PRB}%
  \BibitemOpen
  \bibfield  {author} {\bibinfo {author} {\bibfnamefont {A.~X.}\ \bibnamefont
  {Gray}}, \bibinfo {author} {\bibfnamefont {A.}~\bibnamefont {Janotti}},
  \bibinfo {author} {\bibfnamefont {J.}~\bibnamefont {Son}}, \bibinfo {author}
  {\bibfnamefont {J.~M.}\ \bibnamefont {LeBeau}}, \bibinfo {author}
  {\bibfnamefont {S.}~\bibnamefont {Ueda}}, \bibinfo {author} {\bibfnamefont
  {Y.}~\bibnamefont {Yamashita}}, \bibinfo {author} {\bibfnamefont
  {K.}~\bibnamefont {Kobayashi}}, \bibinfo {author} {\bibfnamefont {A.~M.}\
  \bibnamefont {Kaiser}}, \bibinfo {author} {\bibfnamefont {R.}~\bibnamefont
  {Sutarto}}, \bibinfo {author} {\bibfnamefont {H.}~\bibnamefont {Wadati}},
  \bibinfo {author} {\bibfnamefont {G.~A.}\ \bibnamefont {Sawatzky}}, \bibinfo
  {author} {\bibfnamefont {C.~G.}\ \bibnamefont {Van~de Walle}}, \bibinfo
  {author} {\bibfnamefont {S.}~\bibnamefont {Stemmer}},\ and\ \bibinfo {author}
  {\bibfnamefont {C.~S.}\ \bibnamefont {Fadley}},\ }\bibfield  {title}
  {\bibinfo {title} {Insulating state of ultrathin epitaxial lanio${}_{3}$ thin
  films detected by hard x-ray photoemission},\ }\href
  {https://doi.org/10.1103/PhysRevB.84.075104} {\bibfield  {journal} {\bibinfo
  {journal} {Phys. Rev. B}\ }\textbf {\bibinfo {volume} {84}},\ \bibinfo
  {pages} {075104} (\bibinfo {year} {2011})}\BibitemShut {NoStop}%
\bibitem [{\citenamefont {King}\ \emph {et~al.}(2014)\citenamefont {King},
  \citenamefont {Wei}, \citenamefont {Nie}, \citenamefont {Uchida},
  \citenamefont {Adamo}, \citenamefont {Zhu}, \citenamefont {He}, \citenamefont
  {Bo{\v z}ovi{\'c}}, \citenamefont {Schlom},\ and\ \citenamefont
  {Shen}}]{King2014NNT}%
  \BibitemOpen
  \bibfield  {author} {\bibinfo {author} {\bibfnamefont {P.~D.~C.}\
  \bibnamefont {King}}, \bibinfo {author} {\bibfnamefont {H.~I.}\ \bibnamefont
  {Wei}}, \bibinfo {author} {\bibfnamefont {Y.~F.}\ \bibnamefont {Nie}},
  \bibinfo {author} {\bibfnamefont {M.}~\bibnamefont {Uchida}}, \bibinfo
  {author} {\bibfnamefont {C.}~\bibnamefont {Adamo}}, \bibinfo {author}
  {\bibfnamefont {S.}~\bibnamefont {Zhu}}, \bibinfo {author} {\bibfnamefont
  {X.}~\bibnamefont {He}}, \bibinfo {author} {\bibfnamefont {I.}~\bibnamefont
  {Bo{\v z}ovi{\'c}}}, \bibinfo {author} {\bibfnamefont {D.~G.}\ \bibnamefont
  {Schlom}},\ and\ \bibinfo {author} {\bibfnamefont {K.~M.}\ \bibnamefont
  {Shen}},\ }\bibfield  {title} {\bibinfo {title} {{Atomic-scale control of
  competing electronicphases in ultrathin LaNiO}},\ }\href
  {https://doi.org/10.1038/nnano.2014.59} {\bibfield  {journal} {\bibinfo
  {journal} {Nature Nanotechnology}\ ,\ \bibinfo {pages} {1}} (\bibinfo {year}
  {2014})}\BibitemShut {NoStop}%
\bibitem [{\citenamefont {Golalikhani}\ \emph {et~al.}(2018)\citenamefont
  {Golalikhani}, \citenamefont {Lei}, \citenamefont {Chandrasena},
  \citenamefont {Kasaei}, \citenamefont {Park}, \citenamefont {Bai},
  \citenamefont {Orgiani}, \citenamefont {Ciston}, \citenamefont {Sterbinsky},
  \citenamefont {Arena}, \citenamefont {Shafer}, \citenamefont {Arenholz},
  \citenamefont {Davidson}, \citenamefont {Millis}, \citenamefont {Gray},\ and\
  \citenamefont {Xi}}]{Golalikhani2018}%
  \BibitemOpen
  \bibfield  {author} {\bibinfo {author} {\bibfnamefont {M.}~\bibnamefont
  {Golalikhani}}, \bibinfo {author} {\bibfnamefont {Q.}~\bibnamefont {Lei}},
  \bibinfo {author} {\bibfnamefont {R.~U.}\ \bibnamefont {Chandrasena}},
  \bibinfo {author} {\bibfnamefont {L.}~\bibnamefont {Kasaei}}, \bibinfo
  {author} {\bibfnamefont {H.}~\bibnamefont {Park}}, \bibinfo {author}
  {\bibfnamefont {J.}~\bibnamefont {Bai}}, \bibinfo {author} {\bibfnamefont
  {P.}~\bibnamefont {Orgiani}}, \bibinfo {author} {\bibfnamefont
  {J.}~\bibnamefont {Ciston}}, \bibinfo {author} {\bibfnamefont {G.~E.}\
  \bibnamefont {Sterbinsky}}, \bibinfo {author} {\bibfnamefont {D.~A.}\
  \bibnamefont {Arena}}, \bibinfo {author} {\bibfnamefont {P.}~\bibnamefont
  {Shafer}}, \bibinfo {author} {\bibfnamefont {E.}~\bibnamefont {Arenholz}},
  \bibinfo {author} {\bibfnamefont {B.~A.}\ \bibnamefont {Davidson}}, \bibinfo
  {author} {\bibfnamefont {A.~J.}\ \bibnamefont {Millis}}, \bibinfo {author}
  {\bibfnamefont {A.~X.}\ \bibnamefont {Gray}},\ and\ \bibinfo {author}
  {\bibfnamefont {X.~X.}\ \bibnamefont {Xi}},\ }\bibfield  {title} {\bibinfo
  {title} {{Nature of the metal-insulator transition in few-unit-cell-thick
  LaNiO3 films}},\ }\href {https://doi.org/10.1038/s41467-018-04546-5}
  {\bibfield  {journal} {\bibinfo  {journal} {Nature Communications}\ }\textbf
  {\bibinfo {volume} {9}},\ \bibinfo {pages} {2206} (\bibinfo {year} {2018})},\
  \Eprint {https://arxiv.org/abs/1806.07224} {1806.07224} \BibitemShut
  {NoStop}%
\bibitem [{\citenamefont {Han}\ \emph {et~al.}(2010)\citenamefont {Han},
  \citenamefont {Marianetti},\ and\ \citenamefont {Millis}}]{Han:2010cy}%
  \BibitemOpen
  \bibfield  {author} {\bibinfo {author} {\bibfnamefont {M.~J.}\ \bibnamefont
  {Han}}, \bibinfo {author} {\bibfnamefont {C.~A.}\ \bibnamefont
  {Marianetti}},\ and\ \bibinfo {author} {\bibfnamefont {A.~J.}\ \bibnamefont
  {Millis}},\ }\bibfield  {title} {\bibinfo {title} {{Chemical control of
  orbital polarization in artificially structured transition-metal oxides:
  La$_{2}$Ni$X$O$_{6}$ ($X$=B,Al,Ga,In) from first principles}},\ }\href@noop
  {} {\bibfield  {journal} {\bibinfo  {journal} {Physical Review B}\ }\textbf
  {\bibinfo {volume} {82}},\ \bibinfo {pages} {134408} (\bibinfo {year}
  {2010})}\BibitemShut {NoStop}%
\bibitem [{\citenamefont {Han}\ \emph {et~al.}(2011{\natexlab{a}})\citenamefont
  {Han}, \citenamefont {Wang}, \citenamefont {Marianetti},\ and\ \citenamefont
  {Millis}}]{Han:2011jl}%
  \BibitemOpen
  \bibfield  {author} {\bibinfo {author} {\bibfnamefont {M.~J.}\ \bibnamefont
  {Han}}, \bibinfo {author} {\bibfnamefont {X.}~\bibnamefont {Wang}}, \bibinfo
  {author} {\bibfnamefont {C.~A.}\ \bibnamefont {Marianetti}},\ and\ \bibinfo
  {author} {\bibfnamefont {A.~J.}\ \bibnamefont {Millis}},\ }\bibfield  {title}
  {\bibinfo {title} {{Dynamical Mean-Field Theory of Nickelate
  Superlattices}},\ }\href@noop {} {\bibfield  {journal} {\bibinfo  {journal}
  {Physical Review Letters}\ }\textbf {\bibinfo {volume} {107}},\ \bibinfo
  {pages} {206804} (\bibinfo {year} {2011}{\natexlab{a}})}\BibitemShut
  {NoStop}%
\bibitem [{\citenamefont {Han}\ \emph {et~al.}(2011{\natexlab{b}})\citenamefont
  {Han}, \citenamefont {Wang}, \citenamefont {Marianetti},\ and\ \citenamefont
  {Millis}}]{MJH2011PRL}%
  \BibitemOpen
  \bibfield  {author} {\bibinfo {author} {\bibfnamefont {M.~J.}\ \bibnamefont
  {Han}}, \bibinfo {author} {\bibfnamefont {X.}~\bibnamefont {Wang}}, \bibinfo
  {author} {\bibfnamefont {C.~A.}\ \bibnamefont {Marianetti}},\ and\ \bibinfo
  {author} {\bibfnamefont {A.~J.}\ \bibnamefont {Millis}},\ }\bibfield  {title}
  {\bibinfo {title} {Dynamical mean-field theory of nickelate superlattices},\
  }\href {https://doi.org/10.1103/PhysRevLett.107.206804} {\bibfield  {journal}
  {\bibinfo  {journal} {Phys. Rev. Lett.}\ }\textbf {\bibinfo {volume} {107}},\
  \bibinfo {pages} {206804} (\bibinfo {year} {2011}{\natexlab{b}})}\BibitemShut
  {NoStop}%
\bibitem [{\citenamefont {Han}\ and\ \citenamefont {van
  Veenendaal}(2012)}]{Han:2012bc}%
  \BibitemOpen
  \bibfield  {author} {\bibinfo {author} {\bibfnamefont {M.~J.}\ \bibnamefont
  {Han}}\ and\ \bibinfo {author} {\bibfnamefont {M.}~\bibnamefont {van
  Veenendaal}},\ }\bibfield  {title} {\bibinfo {title} {{Spin-moment formation
  and reduced orbital polarization in LaNiO$_{3}$/LaAlO$_{3}$ superlattice:
  LDA+U study}},\ }\href@noop {} {\bibfield  {journal} {\bibinfo  {journal}
  {Physical Review B}\ }\textbf {\bibinfo {volume} {85}},\ \bibinfo {pages}
  {195102} (\bibinfo {year} {2012})}\BibitemShut {NoStop}%
\bibitem [{\citenamefont {Han}\ \emph {et~al.}(2014)\citenamefont {Han},
  \citenamefont {Kino},\ and\ \citenamefont {Kotani}}]{Han:2014be}%
  \BibitemOpen
  \bibfield  {author} {\bibinfo {author} {\bibfnamefont {M.~J.}\ \bibnamefont
  {Han}}, \bibinfo {author} {\bibfnamefont {H.}~\bibnamefont {Kino}},\ and\
  \bibinfo {author} {\bibfnamefont {T.}~\bibnamefont {Kotani}},\ }\bibfield
  {title} {\bibinfo {title} {{Quasiparticle self-consistent GW study of LaNiO3
  and LaNiO3/LaAlO3 superlattice}},\ }\href@noop {} {\bibfield  {journal}
  {\bibinfo  {journal} {Physical Review B}\ }\textbf {\bibinfo {volume} {90}},\
  \bibinfo {pages} {035127} (\bibinfo {year} {2014})}\BibitemShut {NoStop}%
\bibitem [{\citenamefont {Yoo}\ \emph {et~al.}(2016)\citenamefont {Yoo},
  \citenamefont {Hyun}, \citenamefont {Chang}, \citenamefont {Moreschini},
  \citenamefont {Sohn}, \citenamefont {Kim}, \citenamefont {Bostwick},
  \citenamefont {Rotenberg}, \citenamefont {Shim},\ and\ \citenamefont
  {Noh}}]{Yoo2016PRB}%
  \BibitemOpen
  \bibfield  {author} {\bibinfo {author} {\bibfnamefont {H.~K.}\ \bibnamefont
  {Yoo}}, \bibinfo {author} {\bibfnamefont {S.~I.}\ \bibnamefont {Hyun}},
  \bibinfo {author} {\bibfnamefont {Y.~J.}\ \bibnamefont {Chang}}, \bibinfo
  {author} {\bibfnamefont {L.}~\bibnamefont {Moreschini}}, \bibinfo {author}
  {\bibfnamefont {C.~H.}\ \bibnamefont {Sohn}}, \bibinfo {author}
  {\bibfnamefont {H.-D.}\ \bibnamefont {Kim}}, \bibinfo {author} {\bibfnamefont
  {A.}~\bibnamefont {Bostwick}}, \bibinfo {author} {\bibfnamefont
  {E.}~\bibnamefont {Rotenberg}}, \bibinfo {author} {\bibfnamefont {J.~H.}\
  \bibnamefont {Shim}},\ and\ \bibinfo {author} {\bibfnamefont {T.~W.}\
  \bibnamefont {Noh}},\ }\bibfield  {title} {\bibinfo {title}
  {Thickness-dependent electronic structure in ultrathin ${\mathrm{lanio}}_{3}$
  films under tensile strain},\ }\href
  {https://doi.org/10.1103/PhysRevB.93.035141} {\bibfield  {journal} {\bibinfo
  {journal} {Phys. Rev. B}\ }\textbf {\bibinfo {volume} {93}},\ \bibinfo
  {pages} {035141} (\bibinfo {year} {2016})}\BibitemShut {NoStop}%
\bibitem [{\citenamefont {Paul}\ and\ \citenamefont
  {Birol}(2019)}]{Paul2019_Review}%
  \BibitemOpen
  \bibfield  {author} {\bibinfo {author} {\bibfnamefont {A.}~\bibnamefont
  {Paul}}\ and\ \bibinfo {author} {\bibfnamefont {T.}~\bibnamefont {Birol}},\
  }\bibfield  {title} {\bibinfo {title} {{Applications of DFT + DMFT in
  Materials Science}},\ }\href
  {https://doi.org/10.1146/annurev-matsci-070218-121825} {\bibfield  {journal}
  {\bibinfo  {journal} {Annual Review of Materials Research}\ }\textbf
  {\bibinfo {volume} {49}},\ \bibinfo {pages} {1} (\bibinfo {year} {2019})},\
  \Eprint {https://arxiv.org/abs/1809.09246} {1809.09246} \BibitemShut
  {NoStop}%
\bibitem [{\citenamefont {Puggioni}\ \emph {et~al.}(2012)\citenamefont
  {Puggioni}, \citenamefont {Filippetti},\ and\ \citenamefont
  {Fiorentini}}]{Puggioni2012PRB}%
  \BibitemOpen
  \bibfield  {author} {\bibinfo {author} {\bibfnamefont {D.}~\bibnamefont
  {Puggioni}}, \bibinfo {author} {\bibfnamefont {A.}~\bibnamefont
  {Filippetti}},\ and\ \bibinfo {author} {\bibfnamefont {V.}~\bibnamefont
  {Fiorentini}},\ }\bibfield  {title} {\bibinfo {title} {Ordering and multiple
  phase transitions in ultrathin nickelate superlattices},\ }\href
  {https://doi.org/10.1103/PhysRevB.86.195132} {\bibfield  {journal} {\bibinfo
  {journal} {Phys. Rev. B}\ }\textbf {\bibinfo {volume} {86}},\ \bibinfo
  {pages} {195132} (\bibinfo {year} {2012})}\BibitemShut {NoStop}%
\bibitem [{\citenamefont {Frano}\ \emph {et~al.}(2013)\citenamefont {Frano},
  \citenamefont {Schierle}, \citenamefont {Haverkort}, \citenamefont {Lu},
  \citenamefont {Wu}, \citenamefont {Blanco-Canosa}, \citenamefont {Nwankwo},
  \citenamefont {Boris}, \citenamefont {Wochner}, \citenamefont {Cristiani},
  \citenamefont {Habermeier}, \citenamefont {Logvenov}, \citenamefont {Hinkov},
  \citenamefont {Benckiser}, \citenamefont {Weschke},\ and\ \citenamefont
  {Keimer}}]{Frano2013PRL}%
  \BibitemOpen
  \bibfield  {author} {\bibinfo {author} {\bibfnamefont {A.}~\bibnamefont
  {Frano}}, \bibinfo {author} {\bibfnamefont {E.}~\bibnamefont {Schierle}},
  \bibinfo {author} {\bibfnamefont {M.~W.}\ \bibnamefont {Haverkort}}, \bibinfo
  {author} {\bibfnamefont {Y.}~\bibnamefont {Lu}}, \bibinfo {author}
  {\bibfnamefont {M.}~\bibnamefont {Wu}}, \bibinfo {author} {\bibfnamefont
  {S.}~\bibnamefont {Blanco-Canosa}}, \bibinfo {author} {\bibfnamefont
  {U.}~\bibnamefont {Nwankwo}}, \bibinfo {author} {\bibfnamefont {A.~V.}\
  \bibnamefont {Boris}}, \bibinfo {author} {\bibfnamefont {P.}~\bibnamefont
  {Wochner}}, \bibinfo {author} {\bibfnamefont {G.}~\bibnamefont {Cristiani}},
  \bibinfo {author} {\bibfnamefont {H.~U.}\ \bibnamefont {Habermeier}},
  \bibinfo {author} {\bibfnamefont {G.}~\bibnamefont {Logvenov}}, \bibinfo
  {author} {\bibfnamefont {V.}~\bibnamefont {Hinkov}}, \bibinfo {author}
  {\bibfnamefont {E.}~\bibnamefont {Benckiser}}, \bibinfo {author}
  {\bibfnamefont {E.}~\bibnamefont {Weschke}},\ and\ \bibinfo {author}
  {\bibfnamefont {B.}~\bibnamefont {Keimer}},\ }\bibfield  {title} {\bibinfo
  {title} {Orbital control of noncollinear magnetic order in nickel oxide
  heterostructures},\ }\href {https://doi.org/10.1103/PhysRevLett.111.106804}
  {\bibfield  {journal} {\bibinfo  {journal} {Phys. Rev. Lett.}\ }\textbf
  {\bibinfo {volume} {111}},\ \bibinfo {pages} {106804} (\bibinfo {year}
  {2013})}\BibitemShut {NoStop}%
\bibitem [{\citenamefont {Kaiser}\ \emph {et~al.}(2011)\citenamefont {Kaiser},
  \citenamefont {Gray}, \citenamefont {Conti}, \citenamefont {Son},
  \citenamefont {Greer}, \citenamefont {Perona}, \citenamefont {Rattanachata},
  \citenamefont {Saw}, \citenamefont {Bostwick}, \citenamefont {Yang},
  \citenamefont {Yang}, \citenamefont {Gullikson}, \citenamefont {Kortright},
  \citenamefont {Stemmer},\ and\ \citenamefont {Fadley}}]{Kaiser2011PRL}%
  \BibitemOpen
  \bibfield  {author} {\bibinfo {author} {\bibfnamefont {A.~M.}\ \bibnamefont
  {Kaiser}}, \bibinfo {author} {\bibfnamefont {A.~X.}\ \bibnamefont {Gray}},
  \bibinfo {author} {\bibfnamefont {G.}~\bibnamefont {Conti}}, \bibinfo
  {author} {\bibfnamefont {J.}~\bibnamefont {Son}}, \bibinfo {author}
  {\bibfnamefont {A.}~\bibnamefont {Greer}}, \bibinfo {author} {\bibfnamefont
  {A.}~\bibnamefont {Perona}}, \bibinfo {author} {\bibfnamefont
  {A.}~\bibnamefont {Rattanachata}}, \bibinfo {author} {\bibfnamefont {A.~Y.}\
  \bibnamefont {Saw}}, \bibinfo {author} {\bibfnamefont {A.}~\bibnamefont
  {Bostwick}}, \bibinfo {author} {\bibfnamefont {S.}~\bibnamefont {Yang}},
  \bibinfo {author} {\bibfnamefont {S.-H.}\ \bibnamefont {Yang}}, \bibinfo
  {author} {\bibfnamefont {E.~M.}\ \bibnamefont {Gullikson}}, \bibinfo {author}
  {\bibfnamefont {J.~B.}\ \bibnamefont {Kortright}}, \bibinfo {author}
  {\bibfnamefont {S.}~\bibnamefont {Stemmer}},\ and\ \bibinfo {author}
  {\bibfnamefont {C.~S.}\ \bibnamefont {Fadley}},\ }\bibfield  {title}
  {\bibinfo {title} {Suppression of near-fermi level electronic states at the
  interface in a ${\mathrm{lanio}}_{3}/{\mathrm{srtio}}_{3}$ superlattice},\
  }\href {https://doi.org/10.1103/PhysRevLett.107.116402} {\bibfield  {journal}
  {\bibinfo  {journal} {Phys. Rev. Lett.}\ }\textbf {\bibinfo {volume} {107}},\
  \bibinfo {pages} {116402} (\bibinfo {year} {2011})}\BibitemShut {NoStop}%
\bibitem [{\citenamefont {Tung}\ \emph {et~al.}(2017)\citenamefont {Tung},
  \citenamefont {Luo}, \citenamefont {Lee}, \citenamefont {Chang},
  \citenamefont {Moyer}, \citenamefont {Hong}, \citenamefont {Bedzyk},
  \citenamefont {Zhou}, \citenamefont {Morgan}, \citenamefont {Fong},\ and\
  \citenamefont {Freeland}}]{Tung2017}%
  \BibitemOpen
  \bibfield  {author} {\bibinfo {author} {\bibfnamefont {I.-C.}\ \bibnamefont
  {Tung}}, \bibinfo {author} {\bibfnamefont {G.}~\bibnamefont {Luo}}, \bibinfo
  {author} {\bibfnamefont {J.~H.}\ \bibnamefont {Lee}}, \bibinfo {author}
  {\bibfnamefont {S.~H.}\ \bibnamefont {Chang}}, \bibinfo {author}
  {\bibfnamefont {J.}~\bibnamefont {Moyer}}, \bibinfo {author} {\bibfnamefont
  {H.}~\bibnamefont {Hong}}, \bibinfo {author} {\bibfnamefont {M.~J.}\
  \bibnamefont {Bedzyk}}, \bibinfo {author} {\bibfnamefont {H.}~\bibnamefont
  {Zhou}}, \bibinfo {author} {\bibfnamefont {D.}~\bibnamefont {Morgan}},
  \bibinfo {author} {\bibfnamefont {D.~D.}\ \bibnamefont {Fong}},\ and\
  \bibinfo {author} {\bibfnamefont {J.~W.}\ \bibnamefont {Freeland}},\
  }\bibfield  {title} {\bibinfo {title} {Polarity-driven oxygen vacancy
  formation in ultrathin ${\mathrm{lanio}}_{3}$ films on
  ${\mathrm{srtio}}_{3}$},\ }\href
  {https://doi.org/10.1103/PhysRevMaterials.1.053404} {\bibfield  {journal}
  {\bibinfo  {journal} {Phys. Rev. Mater.}\ }\textbf {\bibinfo {volume} {1}},\
  \bibinfo {pages} {053404} (\bibinfo {year} {2017})}\BibitemShut {NoStop}%
\bibitem [{\citenamefont {Geisler}\ \emph {et~al.}(2017)\citenamefont
  {Geisler}, \citenamefont {Blanca-Romero},\ and\ \citenamefont
  {Pentcheva}}]{Geisler2017}%
  \BibitemOpen
  \bibfield  {author} {\bibinfo {author} {\bibfnamefont {B.}~\bibnamefont
  {Geisler}}, \bibinfo {author} {\bibfnamefont {A.}~\bibnamefont
  {Blanca-Romero}},\ and\ \bibinfo {author} {\bibfnamefont {R.}~\bibnamefont
  {Pentcheva}},\ }\bibfield  {title} {\bibinfo {title} {Design of $n$- and
  $p$-type oxide thermoelectrics in
  ${\mathrm{lanio}}_{3}/{\mathrm{srtio}}_{3}(001)$ superlattices exploiting
  interface polarity},\ }\href {https://doi.org/10.1103/PhysRevB.95.125301}
  {\bibfield  {journal} {\bibinfo  {journal} {Phys. Rev. B}\ }\textbf {\bibinfo
  {volume} {95}},\ \bibinfo {pages} {125301} (\bibinfo {year}
  {2017})}\BibitemShut {NoStop}%
\bibitem [{\citenamefont {Kresse}\ and\ \citenamefont {Hafner}(1993)}]{VASP1}%
  \BibitemOpen
  \bibfield  {author} {\bibinfo {author} {\bibfnamefont {G.}~\bibnamefont
  {Kresse}}\ and\ \bibinfo {author} {\bibfnamefont {J.}~\bibnamefont
  {Hafner}},\ }\bibfield  {title} {\bibinfo {title} {\textit{Ab initio}
  molecular dynamics for liquid metals},\ }\href
  {https://doi.org/10.1103/PhysRevB.47.558} {\bibfield  {journal} {\bibinfo
  {journal} {Phys. Rev. B}\ }\textbf {\bibinfo {volume} {47}},\ \bibinfo
  {pages} {558} (\bibinfo {year} {1993})}\BibitemShut {NoStop}%
\bibitem [{\citenamefont {Kresse}\ and\ \citenamefont
  {Furthm\"uller}(1996)}]{VASP2}%
  \BibitemOpen
  \bibfield  {author} {\bibinfo {author} {\bibfnamefont {G.}~\bibnamefont
  {Kresse}}\ and\ \bibinfo {author} {\bibfnamefont {J.}~\bibnamefont
  {Furthm\"uller}},\ }\bibfield  {title} {\bibinfo {title} {Efficient iterative
  schemes for \textit{ab initio} total-energy calculations using a plane-wave
  basis set},\ }\href {https://doi.org/10.1103/PhysRevB.54.11169} {\bibfield
  {journal} {\bibinfo  {journal} {Phys. Rev. B}\ }\textbf {\bibinfo {volume}
  {54}},\ \bibinfo {pages} {11169} (\bibinfo {year} {1996})}\BibitemShut
  {NoStop}%
\bibitem [{\citenamefont {Dudarev}\ \emph {et~al.}(1998)\citenamefont
  {Dudarev}, \citenamefont {Botton}, \citenamefont {Savrasov}, \citenamefont
  {Humphreys},\ and\ \citenamefont {Sutton}}]{Dudarev1998PRB}%
  \BibitemOpen
  \bibfield  {author} {\bibinfo {author} {\bibfnamefont {S.~L.}\ \bibnamefont
  {Dudarev}}, \bibinfo {author} {\bibfnamefont {G.~A.}\ \bibnamefont {Botton}},
  \bibinfo {author} {\bibfnamefont {S.~Y.}\ \bibnamefont {Savrasov}}, \bibinfo
  {author} {\bibfnamefont {C.~J.}\ \bibnamefont {Humphreys}},\ and\ \bibinfo
  {author} {\bibfnamefont {A.~P.}\ \bibnamefont {Sutton}},\ }\bibfield  {title}
  {\bibinfo {title} {Electron-energy-loss spectra and the structural stability
  of nickel oxide: An lsda+u study},\ }\href
  {https://doi.org/10.1103/PhysRevB.57.1505} {\bibfield  {journal} {\bibinfo
  {journal} {Phys. Rev. B}\ }\textbf {\bibinfo {volume} {57}},\ \bibinfo
  {pages} {1505} (\bibinfo {year} {1998})}\BibitemShut {NoStop}%
\bibitem [{\citenamefont {Chen}\ and\ \citenamefont
  {Millis}(2016{\natexlab{a}})}]{HChen2016}%
  \BibitemOpen
  \bibfield  {author} {\bibinfo {author} {\bibfnamefont {H.}~\bibnamefont
  {Chen}}\ and\ \bibinfo {author} {\bibfnamefont {A.~J.}\ \bibnamefont
  {Millis}},\ }\bibfield  {title} {\bibinfo {title} {Comparative study of
  exchange-correlation functionals for accurate predictions of structural and
  magnetic properties of multiferroic oxides},\ }\href
  {https://doi.org/10.1103/PhysRevB.93.205110} {\bibfield  {journal} {\bibinfo
  {journal} {Phys. Rev. B}\ }\textbf {\bibinfo {volume} {93}},\ \bibinfo
  {pages} {205110} (\bibinfo {year} {2016}{\natexlab{a}})}\BibitemShut
  {NoStop}%
\bibitem [{\citenamefont {Chen}\ and\ \citenamefont
  {Millis}(2016{\natexlab{b}})}]{HChen2016b}%
  \BibitemOpen
  \bibfield  {author} {\bibinfo {author} {\bibfnamefont {H.}~\bibnamefont
  {Chen}}\ and\ \bibinfo {author} {\bibfnamefont {A.~J.}\ \bibnamefont
  {Millis}},\ }\bibfield  {title} {\bibinfo {title} {Spin-density functional
  theories and their $+u$ and $+j$ extensions: A comparative study of
  transition metals and transition metal oxides},\ }\href
  {https://doi.org/10.1103/PhysRevB.93.045133} {\bibfield  {journal} {\bibinfo
  {journal} {Phys. Rev. B}\ }\textbf {\bibinfo {volume} {93}},\ \bibinfo
  {pages} {045133} (\bibinfo {year} {2016}{\natexlab{b}})}\BibitemShut
  {NoStop}%
\bibitem [{\citenamefont {Park}\ \emph
  {et~al.}(2015{\natexlab{a}})\citenamefont {Park}, \citenamefont {Millis},\
  and\ \citenamefont {Marianetti}}]{HPark2015}%
  \BibitemOpen
  \bibfield  {author} {\bibinfo {author} {\bibfnamefont {H.}~\bibnamefont
  {Park}}, \bibinfo {author} {\bibfnamefont {A.~J.}\ \bibnamefont {Millis}},\
  and\ \bibinfo {author} {\bibfnamefont {C.~A.}\ \bibnamefont {Marianetti}},\
  }\bibfield  {title} {\bibinfo {title} {Density functional versus spin-density
  functional and the choice of correlated subspace in multivariable effective
  action theories of electronic structure},\ }\href
  {https://doi.org/10.1103/PhysRevB.92.035146} {\bibfield  {journal} {\bibinfo
  {journal} {Phys. Rev. B}\ }\textbf {\bibinfo {volume} {92}},\ \bibinfo
  {pages} {035146} (\bibinfo {year} {2015}{\natexlab{a}})}\BibitemShut
  {NoStop}%
\bibitem [{\citenamefont {Ryee}\ and\ \citenamefont
  {Han}(2018{\natexlab{a}})}]{Ryee2018b}%
  \BibitemOpen
  \bibfield  {author} {\bibinfo {author} {\bibfnamefont {S.}~\bibnamefont
  {Ryee}}\ and\ \bibinfo {author} {\bibfnamefont {M.~J.}\ \bibnamefont {Han}},\
  }\bibfield  {title} {\bibinfo {title} {The effect of double counting, spin
  density, and hund interaction in the different dft+u functionals},\ }\href
  {https://doi.org/10.1038/s41598-018-27731-4} {\bibfield  {journal} {\bibinfo
  {journal} {Scientific Reports}\ }\textbf {\bibinfo {volume} {8}},\ \bibinfo
  {pages} {9559} (\bibinfo {year} {2018}{\natexlab{a}})}\BibitemShut {NoStop}%
\bibitem [{\citenamefont {Ryee}\ and\ \citenamefont
  {Han}(2018{\natexlab{b}})}]{Ryee2018}%
  \BibitemOpen
  \bibfield  {author} {\bibinfo {author} {\bibfnamefont {S.}~\bibnamefont
  {Ryee}}\ and\ \bibinfo {author} {\bibfnamefont {M.~J.}\ \bibnamefont {Han}},\
  }\bibfield  {title} {\bibinfo {title} {Comparative study of dft+u functionals
  for non-collinear magnetism},\ }\href
  {https://doi.org/10.1088/1361-648X/aac79c} {\bibfield  {journal} {\bibinfo
  {journal} {Journal of Physics: Condensed Matter}\ }\textbf {\bibinfo {volume}
  {30}},\ \bibinfo {pages} {275802} (\bibinfo {year}
  {2018}{\natexlab{b}})}\BibitemShut {NoStop}%
\bibitem [{\citenamefont {Jang}\ \emph {et~al.}(2018)\citenamefont {Jang},
  \citenamefont {Ryee}, \citenamefont {Yoon},\ and\ \citenamefont
  {Han}}]{Jang2018}%
  \BibitemOpen
  \bibfield  {author} {\bibinfo {author} {\bibfnamefont {S.~W.}\ \bibnamefont
  {Jang}}, \bibinfo {author} {\bibfnamefont {S.}~\bibnamefont {Ryee}}, \bibinfo
  {author} {\bibfnamefont {H.}~\bibnamefont {Yoon}},\ and\ \bibinfo {author}
  {\bibfnamefont {M.~J.}\ \bibnamefont {Han}},\ }\bibfield  {title} {\bibinfo
  {title} {Charge density functional plus $u$ theory of ${\mathrm{lamno}}_{3}$:
  Phase diagram, electronic structure, and magnetic interaction},\ }\href
  {https://doi.org/10.1103/PhysRevB.98.125126} {\bibfield  {journal} {\bibinfo
  {journal} {Phys. Rev. B}\ }\textbf {\bibinfo {volume} {98}},\ \bibinfo
  {pages} {125126} (\bibinfo {year} {2018})}\BibitemShut {NoStop}%
\bibitem [{\citenamefont {Jang}\ \emph {et~al.}(2021)\citenamefont {Jang},
  \citenamefont {Kiem}, \citenamefont {Lee}, \citenamefont {Kang},
  \citenamefont {Yoon},\ and\ \citenamefont {Han}}]{Jang2021}%
  \BibitemOpen
  \bibfield  {author} {\bibinfo {author} {\bibfnamefont {S.~W.}\ \bibnamefont
  {Jang}}, \bibinfo {author} {\bibfnamefont {D.~H.}\ \bibnamefont {Kiem}},
  \bibinfo {author} {\bibfnamefont {J.}~\bibnamefont {Lee}}, \bibinfo {author}
  {\bibfnamefont {Y.-G.}\ \bibnamefont {Kang}}, \bibinfo {author}
  {\bibfnamefont {H.}~\bibnamefont {Yoon}},\ and\ \bibinfo {author}
  {\bibfnamefont {M.~J.}\ \bibnamefont {Han}},\ }\bibfield  {title} {\bibinfo
  {title} {Hund's physics and the magnetic ground state of $\mathrm{CrO}x$
  $(x=\text{Cl},\phantom{\rule{0.16em}{0ex}}\text{Br})$},\ }\href
  {https://doi.org/10.1103/PhysRevMaterials.5.034409} {\bibfield  {journal}
  {\bibinfo  {journal} {Phys. Rev. Mater.}\ }\textbf {\bibinfo {volume} {5}},\
  \bibinfo {pages} {034409} (\bibinfo {year} {2021})}\BibitemShut {NoStop}%
\bibitem [{\citenamefont {Lee}\ \emph {et~al.}(2019)\citenamefont {Lee},
  \citenamefont {Jeong}, \citenamefont {Sim}, \citenamefont {Yoon},
  \citenamefont {Ryee},\ and\ \citenamefont {Han}}]{Lee2019}%
  \BibitemOpen
  \bibfield  {author} {\bibinfo {author} {\bibfnamefont {H.}~\bibnamefont
  {Lee}}, \bibinfo {author} {\bibfnamefont {M.~Y.}\ \bibnamefont {Jeong}},
  \bibinfo {author} {\bibfnamefont {J.-H.}\ \bibnamefont {Sim}}, \bibinfo
  {author} {\bibfnamefont {H.}~\bibnamefont {Yoon}}, \bibinfo {author}
  {\bibfnamefont {S.}~\bibnamefont {Ryee}},\ and\ \bibinfo {author}
  {\bibfnamefont {M.~J.}\ \bibnamefont {Han}},\ }\bibfield  {title} {\bibinfo
  {title} {Charge density functional plus u calculation of lacunar spinel
  gam4se8(m = nb, mo, ta, and w)},\ }\href
  {https://doi.org/10.1209/0295-5075/125/47005} {\bibfield  {journal} {\bibinfo
   {journal} {Europhysics Letters}\ }\textbf {\bibinfo {volume} {125}},\
  \bibinfo {pages} {47005} (\bibinfo {year} {2019})}\BibitemShut {NoStop}%
\bibitem [{\citenamefont {May}\ \emph {et~al.}(2010)\citenamefont {May},
  \citenamefont {Kim}, \citenamefont {Rondinelli}, \citenamefont {Karapetrova},
  \citenamefont {Spaldin}, \citenamefont {Bhattacharya},\ and\ \citenamefont
  {Ryan}}]{May2010PRB}%
  \BibitemOpen
  \bibfield  {author} {\bibinfo {author} {\bibfnamefont {S.~J.}\ \bibnamefont
  {May}}, \bibinfo {author} {\bibfnamefont {J.-W.}\ \bibnamefont {Kim}},
  \bibinfo {author} {\bibfnamefont {J.~M.}\ \bibnamefont {Rondinelli}},
  \bibinfo {author} {\bibfnamefont {E.}~\bibnamefont {Karapetrova}}, \bibinfo
  {author} {\bibfnamefont {N.~A.}\ \bibnamefont {Spaldin}}, \bibinfo {author}
  {\bibfnamefont {A.}~\bibnamefont {Bhattacharya}},\ and\ \bibinfo {author}
  {\bibfnamefont {P.~J.}\ \bibnamefont {Ryan}},\ }\bibfield  {title} {\bibinfo
  {title} {Quantifying octahedral rotations in strained perovskite oxide
  films},\ }\href {https://doi.org/10.1103/PhysRevB.82.014110} {\bibfield
  {journal} {\bibinfo  {journal} {Phys. Rev. B}\ }\textbf {\bibinfo {volume}
  {82}},\ \bibinfo {pages} {014110} (\bibinfo {year} {2010})}\BibitemShut
  {NoStop}%
\bibitem [{\citenamefont {Rondinelli}\ and\ \citenamefont
  {Spaldin}(2010)}]{Rondinelli2010PRB}%
  \BibitemOpen
  \bibfield  {author} {\bibinfo {author} {\bibfnamefont {J.~M.}\ \bibnamefont
  {Rondinelli}}\ and\ \bibinfo {author} {\bibfnamefont {N.~A.}\ \bibnamefont
  {Spaldin}},\ }\bibfield  {title} {\bibinfo {title} {Substrate coherency
  driven octahedral rotations in perovskite oxide films},\ }\href
  {https://doi.org/10.1103/PhysRevB.82.113402} {\bibfield  {journal} {\bibinfo
  {journal} {Phys. Rev. B}\ }\textbf {\bibinfo {volume} {82}},\ \bibinfo
  {pages} {113402} (\bibinfo {year} {2010})}\BibitemShut {NoStop}%
\bibitem [{\citenamefont {Hatt}\ and\ \citenamefont
  {Spaldin}(2010)}]{Hatt2010PRB}%
  \BibitemOpen
  \bibfield  {author} {\bibinfo {author} {\bibfnamefont {A.~J.}\ \bibnamefont
  {Hatt}}\ and\ \bibinfo {author} {\bibfnamefont {N.~A.}\ \bibnamefont
  {Spaldin}},\ }\bibfield  {title} {\bibinfo {title} {Structural phases of
  strained ${\text{laalo}}_{3}$ driven by octahedral tilt instabilities},\
  }\href {https://doi.org/10.1103/PhysRevB.82.195402} {\bibfield  {journal}
  {\bibinfo  {journal} {Phys. Rev. B}\ }\textbf {\bibinfo {volume} {82}},\
  \bibinfo {pages} {195402} (\bibinfo {year} {2010})}\BibitemShut {NoStop}%
\bibitem [{\citenamefont {Rondinelli}\ and\ \citenamefont
  {Coh}(2011)}]{Rondinelli2011PRL}%
  \BibitemOpen
  \bibfield  {author} {\bibinfo {author} {\bibfnamefont {J.~M.}\ \bibnamefont
  {Rondinelli}}\ and\ \bibinfo {author} {\bibfnamefont {S.}~\bibnamefont
  {Coh}},\ }\bibfield  {title} {\bibinfo {title} {Large isosymmetric
  reorientation of oxygen octahedra rotation axes in epitaxially strained
  perovskites},\ }\href {https://doi.org/10.1103/PhysRevLett.106.235502}
  {\bibfield  {journal} {\bibinfo  {journal} {Phys. Rev. Lett.}\ }\textbf
  {\bibinfo {volume} {106}},\ \bibinfo {pages} {235502} (\bibinfo {year}
  {2011})}\BibitemShut {NoStop}%
\bibitem [{\citenamefont {Lau}\ and\ \citenamefont
  {Millis}(2013)}]{Bayo2013PRL}%
  \BibitemOpen
  \bibfield  {author} {\bibinfo {author} {\bibfnamefont {B.}~\bibnamefont
  {Lau}}\ and\ \bibinfo {author} {\bibfnamefont {A.~J.}\ \bibnamefont
  {Millis}},\ }\bibfield  {title} {\bibinfo {title} {Theory of the magnetic and
  metal-insulator transitions in $r{\mathrm{nio}}_{3}$ bulk and layered
  structures},\ }\href {https://doi.org/10.1103/PhysRevLett.110.126404}
  {\bibfield  {journal} {\bibinfo  {journal} {Phys. Rev. Lett.}\ }\textbf
  {\bibinfo {volume} {110}},\ \bibinfo {pages} {126404} (\bibinfo {year}
  {2013})}\BibitemShut {NoStop}%
\bibitem [{\citenamefont {Blanca-Romero}\ and\ \citenamefont
  {Pentcheva}(2011)}]{Pentcheva2011PRB}%
  \BibitemOpen
  \bibfield  {author} {\bibinfo {author} {\bibfnamefont {A.}~\bibnamefont
  {Blanca-Romero}}\ and\ \bibinfo {author} {\bibfnamefont {R.}~\bibnamefont
  {Pentcheva}},\ }\bibfield  {title} {\bibinfo {title} {Confinement-induced
  metal-to-insulator transition in strained lanio${}_{3}$/laalo${}_{3}$
  superlattices},\ }\href {https://doi.org/10.1103/PhysRevB.84.195450}
  {\bibfield  {journal} {\bibinfo  {journal} {Phys. Rev. B}\ }\textbf {\bibinfo
  {volume} {84}},\ \bibinfo {pages} {195450} (\bibinfo {year}
  {2011})}\BibitemShut {NoStop}%
\bibitem [{\citenamefont {Park}\ \emph {et~al.}(2012)\citenamefont {Park},
  \citenamefont {Millis},\ and\ \citenamefont {Marianetti}}]{Park2012PRL}%
  \BibitemOpen
  \bibfield  {author} {\bibinfo {author} {\bibfnamefont {H.}~\bibnamefont
  {Park}}, \bibinfo {author} {\bibfnamefont {A.~J.}\ \bibnamefont {Millis}},\
  and\ \bibinfo {author} {\bibfnamefont {C.~A.}\ \bibnamefont {Marianetti}},\
  }\bibfield  {title} {\bibinfo {title} {Site-selective mott transition in
  rare-earth-element nickelates},\ }\href
  {https://doi.org/10.1103/PhysRevLett.109.156402} {\bibfield  {journal}
  {\bibinfo  {journal} {Phys. Rev. Lett.}\ }\textbf {\bibinfo {volume} {109}},\
  \bibinfo {pages} {156402} (\bibinfo {year} {2012})}\BibitemShut {NoStop}%
\bibitem [{\citenamefont {Kim}\ and\ \citenamefont {Han}(2015)}]{HSK2015PRB}%
  \BibitemOpen
  \bibfield  {author} {\bibinfo {author} {\bibfnamefont {H.-S.}\ \bibnamefont
  {Kim}}\ and\ \bibinfo {author} {\bibfnamefont {M.~J.}\ \bibnamefont {Han}},\
  }\bibfield  {title} {\bibinfo {title} {Effect of charge doping on the
  electronic structure, orbital polarization, and structural distortion in
  nickelate superlattice},\ }\href {https://doi.org/10.1103/PhysRevB.91.235102}
  {\bibfield  {journal} {\bibinfo  {journal} {Phys. Rev. B}\ }\textbf {\bibinfo
  {volume} {91}},\ \bibinfo {pages} {235102} (\bibinfo {year}
  {2015})}\BibitemShut {NoStop}%
\bibitem [{\citenamefont {Chaloupka}\ and\ \citenamefont
  {Khaliullin}(2008)}]{Chaloupka2008PRL}%
  \BibitemOpen
  \bibfield  {author} {\bibinfo {author} {\bibfnamefont {J.~c.~v.}\
  \bibnamefont {Chaloupka}}\ and\ \bibinfo {author} {\bibfnamefont
  {G.}~\bibnamefont {Khaliullin}},\ }\bibfield  {title} {\bibinfo {title}
  {Orbital order and possible superconductivity in
  ${\mathrm{lanio}}_{3}/{\mathrm{lamo}}_{3}$ superlattices},\ }\href
  {https://doi.org/10.1103/PhysRevLett.100.016404} {\bibfield  {journal}
  {\bibinfo  {journal} {Phys. Rev. Lett.}\ }\textbf {\bibinfo {volume} {100}},\
  \bibinfo {pages} {016404} (\bibinfo {year} {2008})}\BibitemShut {NoStop}%
\bibitem [{\citenamefont {He}\ and\ \citenamefont
  {Millis}(2015)}]{Zhuoran2015}%
  \BibitemOpen
  \bibfield  {author} {\bibinfo {author} {\bibfnamefont {Z.}~\bibnamefont
  {He}}\ and\ \bibinfo {author} {\bibfnamefont {A.~J.}\ \bibnamefont
  {Millis}},\ }\bibfield  {title} {\bibinfo {title} {Strain control of
  electronic phase in rare-earth nickelates},\ }\href
  {https://doi.org/10.1103/PhysRevB.91.195138} {\bibfield  {journal} {\bibinfo
  {journal} {Phys. Rev. B}\ }\textbf {\bibinfo {volume} {91}},\ \bibinfo
  {pages} {195138} (\bibinfo {year} {2015})}\BibitemShut {NoStop}%
\bibitem [{\citenamefont {Henkelman}\ \emph {et~al.}(2006)\citenamefont
  {Henkelman}, \citenamefont {Arnaldsson},\ and\ \citenamefont
  {J\`{o}nsson}}]{Bader}%
  \BibitemOpen
  \bibfield  {author} {\bibinfo {author} {\bibfnamefont {G.}~\bibnamefont
  {Henkelman}}, \bibinfo {author} {\bibfnamefont {A.}~\bibnamefont
  {Arnaldsson}},\ and\ \bibinfo {author} {\bibfnamefont {H.}~\bibnamefont
  {J\`{o}nsson}},\ }\bibfield  {title} {\bibinfo {title} {A fast and robust
  algorithm for bader decomposition of charge density},\ }\href
  {https://doi.org/http://dx.doi.org/10.1016/j.commatsci.2005.04.010}
  {\bibfield  {journal} {\bibinfo  {journal} {Computational Materials Science}\
  }\textbf {\bibinfo {volume} {36}},\ \bibinfo {pages} {354 } (\bibinfo {year}
  {2006})}\BibitemShut {NoStop}%
\bibitem [{\citenamefont {Mazin}\ \emph {et~al.}(2007)\citenamefont {Mazin},
  \citenamefont {Khomskii}, \citenamefont {Lengsdorf}, \citenamefont {Alonso},
  \citenamefont {Marshall}, \citenamefont {Ibberson}, \citenamefont
  {Podlesnyak}, \citenamefont {Mart\'{\i}nez-Lope},\ and\ \citenamefont
  {Abd-Elmeguid}}]{Mazin2007PRL}%
  \BibitemOpen
  \bibfield  {author} {\bibinfo {author} {\bibfnamefont {I.~I.}\ \bibnamefont
  {Mazin}}, \bibinfo {author} {\bibfnamefont {D.~I.}\ \bibnamefont {Khomskii}},
  \bibinfo {author} {\bibfnamefont {R.}~\bibnamefont {Lengsdorf}}, \bibinfo
  {author} {\bibfnamefont {J.~A.}\ \bibnamefont {Alonso}}, \bibinfo {author}
  {\bibfnamefont {W.~G.}\ \bibnamefont {Marshall}}, \bibinfo {author}
  {\bibfnamefont {R.~M.}\ \bibnamefont {Ibberson}}, \bibinfo {author}
  {\bibfnamefont {A.}~\bibnamefont {Podlesnyak}}, \bibinfo {author}
  {\bibfnamefont {M.~J.}\ \bibnamefont {Mart\'{\i}nez-Lope}},\ and\ \bibinfo
  {author} {\bibfnamefont {M.~M.}\ \bibnamefont {Abd-Elmeguid}},\ }\bibfield
  {title} {\bibinfo {title} {Charge ordering as alternative to jahn-teller
  distortion},\ }\href {https://doi.org/10.1103/PhysRevLett.98.176406}
  {\bibfield  {journal} {\bibinfo  {journal} {Phys. Rev. Lett.}\ }\textbf
  {\bibinfo {volume} {98}},\ \bibinfo {pages} {176406} (\bibinfo {year}
  {2007})}\BibitemShut {NoStop}%
\bibitem [{\citenamefont {Park}\ \emph
  {et~al.}(2015{\natexlab{b}})\citenamefont {Park}, \citenamefont {Millis},\
  and\ \citenamefont {Marianetti}}]{Park2015PRB}%
  \BibitemOpen
  \bibfield  {author} {\bibinfo {author} {\bibfnamefont {H.}~\bibnamefont
  {Park}}, \bibinfo {author} {\bibfnamefont {A.~J.}\ \bibnamefont {Millis}},\
  and\ \bibinfo {author} {\bibfnamefont {C.~A.}\ \bibnamefont {Marianetti}},\
  }\bibfield  {title} {\bibinfo {title} {Density functional versus spin-density
  functional and the choice of correlated subspace in multivariable effective
  action theories of electronic structure},\ }\href
  {https://doi.org/10.1103/PhysRevB.92.035146} {\bibfield  {journal} {\bibinfo
  {journal} {Phys. Rev. B}\ }\textbf {\bibinfo {volume} {92}},\ \bibinfo
  {pages} {035146} (\bibinfo {year} {2015}{\natexlab{b}})}\BibitemShut
  {NoStop}%
\bibitem [{\citenamefont {Mizokawa}\ \emph {et~al.}(2000)\citenamefont
  {Mizokawa}, \citenamefont {Khomskii},\ and\ \citenamefont
  {Sawatzky}}]{Mizokawa2000PRB}%
  \BibitemOpen
  \bibfield  {author} {\bibinfo {author} {\bibfnamefont {T.}~\bibnamefont
  {Mizokawa}}, \bibinfo {author} {\bibfnamefont {D.~I.}\ \bibnamefont
  {Khomskii}},\ and\ \bibinfo {author} {\bibfnamefont {G.~A.}\ \bibnamefont
  {Sawatzky}},\ }\bibfield  {title} {\bibinfo {title} {Spin and charge ordering
  in self-doped mott insulators},\ }\href
  {https://doi.org/10.1103/PhysRevB.61.11263} {\bibfield  {journal} {\bibinfo
  {journal} {Phys. Rev. B}\ }\textbf {\bibinfo {volume} {61}},\ \bibinfo
  {pages} {11263} (\bibinfo {year} {2000})}\BibitemShut {NoStop}%
\bibitem [{\citenamefont {Johnston}\ \emph {et~al.}(2014)\citenamefont
  {Johnston}, \citenamefont {Mukherjee}, \citenamefont {Elfimov}, \citenamefont
  {Berciu},\ and\ \citenamefont {Sawatzky}}]{Johnston2014PRL}%
  \BibitemOpen
  \bibfield  {author} {\bibinfo {author} {\bibfnamefont {S.}~\bibnamefont
  {Johnston}}, \bibinfo {author} {\bibfnamefont {A.}~\bibnamefont {Mukherjee}},
  \bibinfo {author} {\bibfnamefont {I.}~\bibnamefont {Elfimov}}, \bibinfo
  {author} {\bibfnamefont {M.}~\bibnamefont {Berciu}},\ and\ \bibinfo {author}
  {\bibfnamefont {G.~A.}\ \bibnamefont {Sawatzky}},\ }\bibfield  {title}
  {\bibinfo {title} {Charge disproportionation without charge transfer in the
  rare-earth-element nickelates as a possible mechanism for the metal-insulator
  transition},\ }\href {https://doi.org/10.1103/PhysRevLett.112.106404}
  {\bibfield  {journal} {\bibinfo  {journal} {Phys. Rev. Lett.}\ }\textbf
  {\bibinfo {volume} {112}},\ \bibinfo {pages} {106404} (\bibinfo {year}
  {2014})}\BibitemShut {NoStop}%
\bibitem [{\citenamefont {Green}\ \emph {et~al.}(2016)\citenamefont {Green},
  \citenamefont {Haverkort},\ and\ \citenamefont {Sawatzky}}]{Green2016}%
  \BibitemOpen
  \bibfield  {author} {\bibinfo {author} {\bibfnamefont {R.~J.}\ \bibnamefont
  {Green}}, \bibinfo {author} {\bibfnamefont {M.~W.}\ \bibnamefont
  {Haverkort}},\ and\ \bibinfo {author} {\bibfnamefont {G.~A.}\ \bibnamefont
  {Sawatzky}},\ }\bibfield  {title} {\bibinfo {title} {Bond disproportionation
  and dynamical charge fluctuations in the perovskite rare-earth nickelates},\
  }\href {https://doi.org/10.1103/PhysRevB.94.195127} {\bibfield  {journal}
  {\bibinfo  {journal} {Phys. Rev. B}\ }\textbf {\bibinfo {volume} {94}},\
  \bibinfo {pages} {195127} (\bibinfo {year} {2016})}\BibitemShut {NoStop}%
\bibitem [{\citenamefont {Haule}\ and\ \citenamefont
  {Pascut}(2017)}]{Pascut2017}%
  \BibitemOpen
  \bibfield  {author} {\bibinfo {author} {\bibfnamefont {K.}~\bibnamefont
  {Haule}}\ and\ \bibinfo {author} {\bibfnamefont {G.~L.}\ \bibnamefont
  {Pascut}},\ }\bibfield  {title} {\bibinfo {title} {Mott transition and
  magnetism in rare earth nickelates and its fingerprint on the x-ray
  scattering},\ }\href {http://arxiv.org/abs/1703.08196} {\bibfield  {journal}
  {\bibinfo  {journal} {arXiv preprint arXiv:1703.08196}\ } (\bibinfo {year}
  {2017})}\BibitemShut {NoStop}%
\bibitem [{\citenamefont {Subedi}\ \emph {et~al.}(2015)\citenamefont {Subedi},
  \citenamefont {Peil},\ and\ \citenamefont {Georges}}]{Subedi2015}%
  \BibitemOpen
  \bibfield  {author} {\bibinfo {author} {\bibfnamefont {A.}~\bibnamefont
  {Subedi}}, \bibinfo {author} {\bibfnamefont {O.~E.}\ \bibnamefont {Peil}},\
  and\ \bibinfo {author} {\bibfnamefont {A.}~\bibnamefont {Georges}},\
  }\bibfield  {title} {\bibinfo {title} {Low-energy description of the
  metal-insulator transition in the rare-earth nickelates},\ }\href
  {https://doi.org/10.1103/PhysRevB.91.075128} {\bibfield  {journal} {\bibinfo
  {journal} {Phys. Rev. B}\ }\textbf {\bibinfo {volume} {91}},\ \bibinfo
  {pages} {075128} (\bibinfo {year} {2015})}\BibitemShut {NoStop}%
\bibitem [{\citenamefont {Khaliullin}(2005)}]{Khaliullin2005PTPS}%
  \BibitemOpen
  \bibfield  {author} {\bibinfo {author} {\bibfnamefont {G.}~\bibnamefont
  {Khaliullin}},\ }\bibfield  {title} {\bibinfo {title} {Orbital order and
  fluctuations in mott insulators},\ }\href
  {https://doi.org/10.1143/PTPS.160.155} {\bibfield  {journal} {\bibinfo
  {journal} {Progress of Theoretical Physics Supplement}\ }\textbf {\bibinfo
  {volume} {160}},\ \bibinfo {pages} {155} (\bibinfo {year}
  {2005})}\BibitemShut {NoStop}%
\bibitem [{\citenamefont {Wu}\ \emph {et~al.}(2013)\citenamefont {Wu},
  \citenamefont {Benckiser}, \citenamefont {Haverkort}, \citenamefont {Frano},
  \citenamefont {Lu}, \citenamefont {Nwankwo}, \citenamefont {Br\"uck},
  \citenamefont {Audehm}, \citenamefont {Goering}, \citenamefont {Macke},
  \citenamefont {Hinkov}, \citenamefont {Wochner}, \citenamefont {Christiani},
  \citenamefont {Heinze}, \citenamefont {Logvenov}, \citenamefont
  {Habermeier},\ and\ \citenamefont {Keimer}}]{Wu2013}%
  \BibitemOpen
  \bibfield  {author} {\bibinfo {author} {\bibfnamefont {M.}~\bibnamefont
  {Wu}}, \bibinfo {author} {\bibfnamefont {E.}~\bibnamefont {Benckiser}},
  \bibinfo {author} {\bibfnamefont {M.~W.}\ \bibnamefont {Haverkort}}, \bibinfo
  {author} {\bibfnamefont {A.}~\bibnamefont {Frano}}, \bibinfo {author}
  {\bibfnamefont {Y.}~\bibnamefont {Lu}}, \bibinfo {author} {\bibfnamefont
  {U.}~\bibnamefont {Nwankwo}}, \bibinfo {author} {\bibfnamefont
  {S.}~\bibnamefont {Br\"uck}}, \bibinfo {author} {\bibfnamefont
  {P.}~\bibnamefont {Audehm}}, \bibinfo {author} {\bibfnamefont
  {E.}~\bibnamefont {Goering}}, \bibinfo {author} {\bibfnamefont
  {S.}~\bibnamefont {Macke}}, \bibinfo {author} {\bibfnamefont
  {V.}~\bibnamefont {Hinkov}}, \bibinfo {author} {\bibfnamefont
  {P.}~\bibnamefont {Wochner}}, \bibinfo {author} {\bibfnamefont
  {G.}~\bibnamefont {Christiani}}, \bibinfo {author} {\bibfnamefont
  {S.}~\bibnamefont {Heinze}}, \bibinfo {author} {\bibfnamefont
  {G.}~\bibnamefont {Logvenov}}, \bibinfo {author} {\bibfnamefont {H.-U.}\
  \bibnamefont {Habermeier}},\ and\ \bibinfo {author} {\bibfnamefont
  {B.}~\bibnamefont {Keimer}},\ }\bibfield  {title} {\bibinfo {title} {Strain
  and composition dependence of orbital polarization in nickel oxide
  superlattices},\ }\href {https://doi.org/10.1103/PhysRevB.88.125124}
  {\bibfield  {journal} {\bibinfo  {journal} {Phys. Rev. B}\ }\textbf {\bibinfo
  {volume} {88}},\ \bibinfo {pages} {125124} (\bibinfo {year}
  {2013})}\BibitemShut {NoStop}%
\bibitem [{\citenamefont {Peil}\ \emph {et~al.}(2014)\citenamefont {Peil},
  \citenamefont {Ferrero},\ and\ \citenamefont {Georges}}]{Peil2014PRB}%
  \BibitemOpen
  \bibfield  {author} {\bibinfo {author} {\bibfnamefont {O.~E.}\ \bibnamefont
  {Peil}}, \bibinfo {author} {\bibfnamefont {M.}~\bibnamefont {Ferrero}},\ and\
  \bibinfo {author} {\bibfnamefont {A.}~\bibnamefont {Georges}},\ }\bibfield
  {title} {\bibinfo {title} {Orbital polarization in strained
  ${\mathrm{lanio}}_{3}$: Structural distortions and correlation effects},\
  }\href {https://doi.org/10.1103/PhysRevB.90.045128} {\bibfield  {journal}
  {\bibinfo  {journal} {Phys. Rev. B}\ }\textbf {\bibinfo {volume} {90}},\
  \bibinfo {pages} {045128} (\bibinfo {year} {2014})}\BibitemShut {NoStop}%
\bibitem [{\citenamefont {Park}\ \emph {et~al.}(2016)\citenamefont {Park},
  \citenamefont {Millis},\ and\ \citenamefont {Marianetti}}]{Park2016PRB}%
  \BibitemOpen
  \bibfield  {author} {\bibinfo {author} {\bibfnamefont {H.}~\bibnamefont
  {Park}}, \bibinfo {author} {\bibfnamefont {A.~J.}\ \bibnamefont {Millis}},\
  and\ \bibinfo {author} {\bibfnamefont {C.~A.}\ \bibnamefont {Marianetti}},\
  }\bibfield  {title} {\bibinfo {title} {Influence of quantum confinement and
  strain on orbital polarization of four-layer ${\mathrm{lanio}}_{3}$
  superlattices: A dft+dmft study},\ }\href
  {https://doi.org/10.1103/PhysRevB.93.235109} {\bibfield  {journal} {\bibinfo
  {journal} {Phys. Rev. B}\ }\textbf {\bibinfo {volume} {93}},\ \bibinfo
  {pages} {235109} (\bibinfo {year} {2016})}\BibitemShut {NoStop}%
\bibitem [{\citenamefont {Chakhalian}\ \emph {et~al.}(2011)\citenamefont
  {Chakhalian}, \citenamefont {Rondinelli}, \citenamefont {Liu}, \citenamefont
  {Gray}, \citenamefont {Kareev}, \citenamefont {Moon}, \citenamefont {Prasai},
  \citenamefont {Cohn}, \citenamefont {Varela}, \citenamefont {Tung},
  \citenamefont {Bedzyk}, \citenamefont {Altendorf}, \citenamefont {Strigari},
  \citenamefont {Dabrowski}, \citenamefont {Tjeng}, \citenamefont {Ryan},\ and\
  \citenamefont {Freeland}}]{Chakhalian:2011ef}%
  \BibitemOpen
  \bibfield  {author} {\bibinfo {author} {\bibfnamefont {J.}~\bibnamefont
  {Chakhalian}}, \bibinfo {author} {\bibfnamefont {J.~M.}\ \bibnamefont
  {Rondinelli}}, \bibinfo {author} {\bibfnamefont {J.}~\bibnamefont {Liu}},
  \bibinfo {author} {\bibfnamefont {B.~A.}\ \bibnamefont {Gray}}, \bibinfo
  {author} {\bibfnamefont {M.}~\bibnamefont {Kareev}}, \bibinfo {author}
  {\bibfnamefont {E.~J.}\ \bibnamefont {Moon}}, \bibinfo {author}
  {\bibfnamefont {N.}~\bibnamefont {Prasai}}, \bibinfo {author} {\bibfnamefont
  {J.~L.}\ \bibnamefont {Cohn}}, \bibinfo {author} {\bibfnamefont
  {M.}~\bibnamefont {Varela}}, \bibinfo {author} {\bibfnamefont {I.~C.}\
  \bibnamefont {Tung}}, \bibinfo {author} {\bibfnamefont {M.~J.}\ \bibnamefont
  {Bedzyk}}, \bibinfo {author} {\bibfnamefont {S.~G.}\ \bibnamefont
  {Altendorf}}, \bibinfo {author} {\bibfnamefont {F.}~\bibnamefont {Strigari}},
  \bibinfo {author} {\bibfnamefont {B.}~\bibnamefont {Dabrowski}}, \bibinfo
  {author} {\bibfnamefont {L.~H.}\ \bibnamefont {Tjeng}}, \bibinfo {author}
  {\bibfnamefont {P.~J.}\ \bibnamefont {Ryan}},\ and\ \bibinfo {author}
  {\bibfnamefont {J.~W.}\ \bibnamefont {Freeland}},\ }\bibfield  {title}
  {\bibinfo {title} {{Asymmetric Orbital-Lattice Interactions in Ultrathin
  Correlated Oxide Films}},\ }\href
  {https://doi.org/10.1103/PhysRevLett.107.116805} {\bibfield  {journal}
  {\bibinfo  {journal} {Physical Review Letters}\ }\textbf {\bibinfo {volume}
  {107}},\ \bibinfo {pages} {116805} (\bibinfo {year} {2011})}\BibitemShut
  {NoStop}%
\bibitem [{\citenamefont {Freeland}\ \emph {et~al.}(2011)\citenamefont
  {Freeland}, \citenamefont {Liu}, \citenamefont {Kareev}, \citenamefont
  {Gray}, \citenamefont {Kim}, \citenamefont {Ryan}, \citenamefont
  {Pentcheva},\ and\ \citenamefont {Chakhalian}}]{Freeland2011}%
  \BibitemOpen
  \bibfield  {author} {\bibinfo {author} {\bibfnamefont {J.~W.}\ \bibnamefont
  {Freeland}}, \bibinfo {author} {\bibfnamefont {J.}~\bibnamefont {Liu}},
  \bibinfo {author} {\bibfnamefont {M.}~\bibnamefont {Kareev}}, \bibinfo
  {author} {\bibfnamefont {B.}~\bibnamefont {Gray}}, \bibinfo {author}
  {\bibfnamefont {J.~W.}\ \bibnamefont {Kim}}, \bibinfo {author} {\bibfnamefont
  {P.}~\bibnamefont {Ryan}}, \bibinfo {author} {\bibfnamefont {R.}~\bibnamefont
  {Pentcheva}},\ and\ \bibinfo {author} {\bibfnamefont {J.}~\bibnamefont
  {Chakhalian}},\ }\bibfield  {title} {\bibinfo {title} {Orbital control in
  strained ultra-thin lanio 3 /laalo 3 superlattices},\ }\href
  {http://stacks.iop.org/0295-5075/96/i=5/a=57004} {\bibfield  {journal}
  {\bibinfo  {journal} {EPL (Europhysics Letters)}\ }\textbf {\bibinfo {volume}
  {96}},\ \bibinfo {pages} {57004} (\bibinfo {year} {2011})}\BibitemShut
  {NoStop}%
\bibitem [{\citenamefont {Wu}\ \emph {et~al.}(2015)\citenamefont {Wu},
  \citenamefont {Benckiser}, \citenamefont {Audehm}, \citenamefont {Goering},
  \citenamefont {Wochner}, \citenamefont {Christiani}, \citenamefont
  {Logvenov}, \citenamefont {Habermeier},\ and\ \citenamefont
  {Keimer}}]{Wu2015PRB}%
  \BibitemOpen
  \bibfield  {author} {\bibinfo {author} {\bibfnamefont {M.}~\bibnamefont
  {Wu}}, \bibinfo {author} {\bibfnamefont {E.}~\bibnamefont {Benckiser}},
  \bibinfo {author} {\bibfnamefont {P.}~\bibnamefont {Audehm}}, \bibinfo
  {author} {\bibfnamefont {E.}~\bibnamefont {Goering}}, \bibinfo {author}
  {\bibfnamefont {P.}~\bibnamefont {Wochner}}, \bibinfo {author} {\bibfnamefont
  {G.}~\bibnamefont {Christiani}}, \bibinfo {author} {\bibfnamefont
  {G.}~\bibnamefont {Logvenov}}, \bibinfo {author} {\bibfnamefont {H.-U.}\
  \bibnamefont {Habermeier}},\ and\ \bibinfo {author} {\bibfnamefont
  {B.}~\bibnamefont {Keimer}},\ }\bibfield  {title} {\bibinfo {title} {Orbital
  reflectometry of ${\text{prnio}}_{3}/{\text{pralo}}_{3}$ superlattices},\
  }\href {https://doi.org/10.1103/PhysRevB.91.195130} {\bibfield  {journal}
  {\bibinfo  {journal} {Phys. Rev. B}\ }\textbf {\bibinfo {volume} {91}},\
  \bibinfo {pages} {195130} (\bibinfo {year} {2015})}\BibitemShut {NoStop}%
\bibitem [{\citenamefont {Lee}\ \emph {et~al.}(2011{\natexlab{a}})\citenamefont
  {Lee}, \citenamefont {Chen},\ and\ \citenamefont {Balents}}]{SBLee2011PRL}%
  \BibitemOpen
  \bibfield  {author} {\bibinfo {author} {\bibfnamefont {S.}~\bibnamefont
  {Lee}}, \bibinfo {author} {\bibfnamefont {R.}~\bibnamefont {Chen}},\ and\
  \bibinfo {author} {\bibfnamefont {L.}~\bibnamefont {Balents}},\ }\bibfield
  {title} {\bibinfo {title} {Landau theory of charge and spin ordering in the
  nickelates},\ }\href {https://doi.org/10.1103/PhysRevLett.106.016405}
  {\bibfield  {journal} {\bibinfo  {journal} {Phys. Rev. Lett.}\ }\textbf
  {\bibinfo {volume} {106}},\ \bibinfo {pages} {016405} (\bibinfo {year}
  {2011}{\natexlab{a}})}\BibitemShut {NoStop}%
\bibitem [{\citenamefont {Lee}\ \emph {et~al.}(2011{\natexlab{b}})\citenamefont
  {Lee}, \citenamefont {Chen},\ and\ \citenamefont {Balents}}]{SBLee2011PRB}%
  \BibitemOpen
  \bibfield  {author} {\bibinfo {author} {\bibfnamefont {S.}~\bibnamefont
  {Lee}}, \bibinfo {author} {\bibfnamefont {R.}~\bibnamefont {Chen}},\ and\
  \bibinfo {author} {\bibfnamefont {L.}~\bibnamefont {Balents}},\ }\bibfield
  {title} {\bibinfo {title} {Metal-insulator transition in a two-band model for
  the perovskite nickelates},\ }\href
  {https://doi.org/10.1103/PhysRevB.84.165119} {\bibfield  {journal} {\bibinfo
  {journal} {Phys. Rev. B}\ }\textbf {\bibinfo {volume} {84}},\ \bibinfo
  {pages} {165119} (\bibinfo {year} {2011}{\natexlab{b}})}\BibitemShut
  {NoStop}%
\end{thebibliography}%

\end{document}